%% file: main.tex
\documentclass[format = sigconf, review=false, balance=false ]{acmart} %{sig-alternate}

\usepackage{algpseudocode}

\usepackage{physics}
\usepackage{xcolor, wrapfig}
\usepackage{soul}
\usepackage[disable]{todonotes}
\usepackage{todonotes}
\usepackage[english]{babel}
\usepackage[utf8]{inputenc}
\usepackage{multirow}
\usepackage{mathtools}
\usepackage{textcomp}
\usepackage{blindtext}
\usepackage{makecell}
\usepackage{enumitem}
\usepackage{algorithm} 
\usepackage{titlesec}
\usepackage{lipsum}
\usepackage[title]{appendix}
\usepackage{acmart-taps}

\def\BibTeX{{\rm B\kern-.05em{\sc i\kern-.025em b}\kern-.08em
    T\kern-.1667em\lower.7ex\hbox{E}\kern-.125emX}}

\let\oldtodo\todo
\renewcommand{\todo}[1]{\oldtodo[inline, color=yellow]{{\bf TODO:} #1}}

\newcommand{\donotshow}[1]{}

\def\eg{{\em e.g.}, }
\def\ie{{\em i.e.}, }

\setlist[enumerate]{topsep=0pt,itemsep=-1ex,partopsep=1ex,parsep=1ex}

\copyrightyear{2023}
\acmYear{2023}
\setcopyright{rightsretained}
\acmConference[MICRO '23]{56th Annual IEEE/ACM International Symposium on
Microarchitecture}{October 28-November 1, 2023}{Toronto, ON, Canada}
\acmBooktitle{56th Annual IEEE/ACM International Symposium on
Microarchitecture (MICRO '23), October 28-November 1, 2023, Toronto, ON,
Canada}
\acmDOI{10.1145/3613424.3614315}
\acmISBN{979-8-4007-0329-4/23/10}
\title{Supporting Energy-Based Learning with an Ising Machine Substrate: A Case Study
on RBM}
\author{Uday Kumar Reddy Vengalam}
\additionalaffiliation{\institution{AMD Research} \country{USA}}
\orcid{0000-0001-6748-0625}
\email{uvengala@ur.rochester.edu}

\affiliation{ \institution{University of Rochester} \country{USA}}

\author {Yongchao Liu} 
\orcid{0009-0005-6756-7911}
\email{yongchao.liu@rochester.edu}
\affiliation{ \institution{University of Rochester} \country{USA}}

\author{Tong Geng} 
\orcid{0000-0002-3644-2922}
\email{tong.geng@rochester.edu}
\affiliation{ \institution{University of Rochester} \country{USA}}

\author{Hui Wu}
\orcid{0009-0008-9150-5608}
\email{hui.wu@rochester.edu}
\affiliation{ \institution{University of Rochester} \country{USA}}

\author{Michael Huang}
\orcid{0000-0001-9799-2920}
\email{michael.huang@rochester.edu}
\affiliation{ \institution{University of Rochester} \country{USA}}

\begin{document}

\begin{abstract}
Nature apparently does a lot of computation constantly. If we can harness some of that computation at an appropriate level, we can potentially perform
certain type of computation (much) faster and more efficiently than we
can do with a von Neumann computer. Indeed, many powerful algorithms are inspired by nature and are thus prime candidates for nature-based computation. One
particular branch of this effort that has seen some recent rapid advances is Ising machines. Some Ising machines are already showing better performance and energy efficiency for optimization problems. Through design iterations
and co-evolution between hardware and algorithm, we expect more benefits from nature-based computing systems in the future. In this paper, we make a case for an augmented Ising machine suitable for both training and inference using an
energy-based machine learning algorithm. We show that with a small change, the Ising substrate accelerates key parts of the algorithm and achieves non-trivial speedup and efficiency gain. With a more substantial change,
we can turn the machine into a self-sufficient gradient follower to virtually complete training entirely in hardware. This can bring about 29x speedup and about 1000x reduction in energy compared to a Tensor Processing
Unit (TPU) host.
\end{abstract}

\maketitle

\input{1intro.tex}

\input{2related.tex}
\input{3arch.tex}
\input{5eval.tex}
\input{6conclusions.tex}

 \begin{acks}
   The authors also thank the anonymous referees for
   their valuable comments and helpful suggestions. This work is
   supported by NSF under Awards No. 2231036
   and 2233378.
\end{acks}

\begin{appendices}
\input{7appendix.tex}

\input{4circuit.tex}
\end{appendices}
\bibliographystyle{ACM-Reference-Format}
\bibliography{main}
\end{document}

%% file: 1intro.tex
\section{Introduction}\label{sec:introduction}

Nature apparently does a lot of computation all the time, solving differential
equations, performing random sampling, and so on. We have harnessed some of it
of course. The transistors, for example, can be turned on and off based on the
law of nature. They are the foundation for most of our computers
today. But this is different from harnessing nature's computational capability
at some higher level, for example, to solve an entire problem. Indeed, some
very powerful algorithms are inspired by nature~\cite{zames1981genetic,Kirkpatrick671,
ackley1985learning}. It is increasingly evident that certain
nature-based systems with fast dynamics can solve a certain set of problems
much more quickly and efficiently than through mapping it to the von Neumann
interface~\cite{afoakwa.hpca21}. 

Among various critical real-world problems, Machine Learning (ML) training is known to be extremely computationally expensive\footnote{ChatGPT is estimated to have cost over 10,000 GPUs in the training and over \$4 million in each training session.}, which calls for a breakthrough solution to reduce the computational time and energy cost. We believe that a nature-based computing substrate could provide a potential fundamental solution to this issue. This idea is actually not entirely new -- recently, Hinton has discussed a similar perspective in \cite{hinton2022forward}. Specifically, he emphasized the significance of utilizing the nature of electronic components (such as capacitors, resistors, and conductors) in analog hardware to enable efficient ML with "mortal computation". However, he also acknowledged that implementing backpropagation using nature-based computing substrates, as opposed to forward propagation, is a significant challenge. 

This motivates us to investigate whether the recent emerging high-performance nature-based computing substrates have the potential for efficient ML.
One particular branch of such computing substrates that has seen some recent rapid advances is Ising machines. In a nutshell, Ising machines leverage nature to seek low energy states for
a system of coupled spins. A number of problems (in fact, all NP-complete
problems) can be expressed as an equivalent optimization problem of the
Ising formula (more on that in Sec.~\ref{sec:background}). Though existing
Ising machines are largely in the form of prototypes and concepts, they are
already showing promise of (much) better performance and energy efficiency for
optimization problems. But the real appeal is in their future opportunities.
First, through design iterations, their computational capacity and
efficiency will continue to improve, hopefully quickly and significantly as
they have in the recent past. Second, with novel hardware, the design of
algorithms (especially those inspired by nature) will co-evolve with the
hardware and lead to a richer combination of problem-solving modalities.

In this paper, we make a concrete case for this second point by demonstrating that an
augmented Ising machine design can bring $29\times$ speedup with $1000\times$ energy cost reduction over TPUs to the training of Restricted Boltzmann Machine (RBM) and variants, a machine learning modality that has drawn considerable attention recently. 
It is worth noting that Boltzmann-Machine-based ML has emerged as a promising approach for solving various real-world networking problems including recommendation systems \cite{liu2023Ising} and traffic prediction \cite{pan2023Ising}, outperforming traditional neural network solutions in terms of both inference speed and accuracy. However, the training of Boltzmann-Machine-based models remains prohibitively expensive, thus hindering the realization of their full potential in the future. The proposed training approach based on (augmented) Ising machines will fundamentally address this problem.

It is also important to note that we believe the true appeal lies in what we can
do when we \emph{co-design} the hardware and the algorithm\footnote{Indeed,
having different substrates can create the same type of cross-pollination
that led to quantum-inspired algorithms~\cite{542334}.}. In this paper, we
mostly follow the existing practice in terms of the training algorithm as this
is merely the first step of our exploration. In the future, we will investigate how to extend the proposed solution to general ML.

%% file: 2related.tex
\section{Background and Related work}\label{sec:background}

\subsection{Principles of Ising machines}

The Ising model is used to describe the Hamiltonian of a system of coupled
spins. The spins have one degree of freedom and take one of two values ($+1$,
$-1$). The energy of the system is a function of pair-wise coupling of the
spins ($J_{ij}$) and the interaction ($h_i$)
of some external field ($\mu$) with each
spin. The resulting Hamiltonian is as follows:
\begin{equation}
\label{eqn:Ising_w_field}
H = -\sum_{{(i<j)}} J_{ij}\sigma_i\sigma_j - \mu \sum_{{i}} h_i\sigma_i
\end{equation}

A physical system with such a Hamiltonian naturally tends towards low-energy
states. It is as if nature tries to solve an optimization problem with
Eq.~\ref{eqn:Ising_w_field} as the objective function, which is not a trivial
task. Indeed, the cardinality of the state space grows exponentially with the
number of spins, and the optimization problem is NP-complete: it is easily
convertible to and from a generalized max-cut problem, which is part of the
original list of NP-complete problem~\cite{Karp1972}.

Thus if a physical system of spins somehow offers programmable coupling
parameters ($J_{ij}$ and $\mu h_i$ in Eq.~\ref{eqn:Ising_w_field}), they can
be used as a special purpose computer to solve\footnote{Throughout
the paper, by ``solving" an optimization problem we mean the
searching for a good solution rather than finding the global optimum, 
equivalent to reaching the ground state.} optimization problems that
can be expressed in Ising formula (Eq.~\ref{eqn:Ising_w_field}). In fact, all
problems in the Karp NP-complete set have their Ising formula
derived~\cite{lucas2014ising}. Also, if a problem already has a QUBO (quadratic
unconstrained binary optimization) formulation, mapping to Ising formula is as
easy as substituting bits for spins: $\sigma_i = 2b_i-1$.

Because of the broad class of problems that can map to the Ising formula,
building nature-based computing systems that solve these problems has
attracted significant attention~\cite{kim2010quantum,bunyk2014architectural,yamaoka201520k,barends2016digitized,berloff2017realizing,king2018observation,bohm2019poor,hamerly2019towards,pierangeli2019large}.

%Loosely speaking, an Ising machine's design goes through four
%steps:
%\fixme{Delete this list if space is limited}
%\begin{enumerate}[nolistsep]
%\setlength{\itemsep}{0cm}
%\item Identify the physical variable to represent a spin (be it a 
%qubit~\cite{PhysRevB.82.024511}, the phase of an optical pulse~\cite{inagaki2016coherent}, 
%or the polarity of a capacitor's voltage~\cite{afoakwa.hpca21});

%\item Identify the mechanism of coupling and how to program the coefficients;

%\item Demonstrate the problem solving capability showing both the theory of its operation (reveal the ``invisible hand" of nature) and satisfactory results of practice;

%\item Demonstrate superior machine metrics (solution time, energy consumption,
%and construction costs).
%\end{enumerate} 

It is important to note that different approaches may
offer different fundamental tradeoffs and goes through varying gestation
speeds. Thus it could be premature to evaluate a general approach based on
observed instances of prototypes. Nevertheless, we provide a broad-brushed
characterization, which can help researchers get a 
basic sense of the landscape -- as long as the caveats are properly understood.

%\begin{comment}
    
\subsection{The three generations of Ising machines}

Some of the earliest and perhaps the best known Ising machines are the
quantum annealers marketed by D-Wave. At the moment, quantum annealers
are extremely sensitive to noise, necessitating cryogenic
operating condition that consumes much power (25KW for
D-Wave 2000q). Also, they use a local coupling
network, which means the the nominal
2000 nodes on the D-Wave 2000q is equivalent to only about 64 effective
nodes~\cite{hamerly2018scaling,hamerly2019experimental}.

Coherent Ising Machines (CIM) can be thought of as a
second-generation design where some of the issues are
addressed~\cite{inagaki2016coherent, McMahon614, 16bit_cim,yamamoto2017coherent,
ncomm_bohm2019}. However, besides their own technical challenges, the
current CIMs all use computed rather than physical coupling. Such a design
will unlikely to be energy-efficient due to fundamental reasons.

Since the operating principle of CIM can be viewed with a 
model that describes coupled oscillators~\cite{Takeda_2017}, 
using other oscillators can in theory achieve a
similar goal. This led to a number of electronic Oscillator-based Ising
Machines (OIM) which can be considered as a third-generation. These
systems use electronic oscillators (LC tanks)~\cite{wang2017oscillator} for 
spins and (programmable) resistors as coupling
units. However, for on-chip integration, inductors are often a source of practical
challenges. They are area intensive, have undesirable parasitics with
reduced quality factor and increased phase noise all of which pose practical
challenges in maintaining frequency uniformity and phase synchronicity
between thousands of on-chip oscillators. 

Another electronic design with a different architecture is the Bistable
Resistively-coupled Ising Machine (BRIM)~\cite{afoakwa.hpca21}. In BRIM, the
spin is implemented as capacitor voltage controlled by a feedback
circuit, making it bistable. The design is CMOS-compatible and since it uses
voltage (as opposed to phase) to represent spin, it enables a straightforward
interface to additional architectural support for computational tasks. We
therefore use a baseline substrate similar to BRIM. Note that the same
principle discussed in the paper could directly apply to all Ising machines
with different amount of glue logic.

%\end{comment}

\subsection{Energy-based models}

The concept of energy is used not only in traditional optimization
algorithms~\cite{Kirkpatrick671} but also in a number of machine
learning algorithms collectively referred to as Energy-Based Models
(EBM)~\cite{lecun2007energy}. The system usually consists of two
sets of variables $X$ and $Y$ (as a concrete example, let us imagine for now
$X$ represents pixels of an image, and $Y$ Boolean variables classifying the
image). If the energy of the state, $E(X,Y)$, is low, then the classification
is good. In many EBMs, the energy is similar to the Ising formula.
In the well-known model of Boltzmann machine~\cite{hinton1984boltzmann} 
for example, if we ignore the distinction between the two set of
variables and just refer to each variable as $\sigma_i$, the energy is
equivalent to the Ising model:\begin{equation}
E = -\sum_{i<j} W_{ij}\sigma_i \sigma_j - \sum_i b_i\sigma_i
\label{eqn:bm}
\end{equation}
When using Boltzmann machines for inference, the system is also similar to
using an Ising machine, but with a difference in the  
the meaning of the variables/spins. But by and large, Ising machines
can accelerate inference of Boltzmann machines in a straightforward
manner.

Training a Boltzmann machine is an entirely different matter.
Unlike in an optimization problem where the weights are inputs,
in training the weights are the output. In other words, training
is the inverse of an combinatoric optimization problem.
Like in many machine learning algorithms, training is done by using a
gradient descent approach to lower the loss function while iterating over
a set of training samples. While we will get into the details later, a key
point to emphasize here is that the key challenge in such a gradient descent
algorithm often involves terms that are computationally intractable, leading
to the necessity of approximation algorithms. Here, a nature-based computing
substrate allows us to adopt approaches convenient or efficient for that
substrate without the need to follow exactly the prevailing von Neumann
algorithms.

In this paper, we will show how a physical Ising machine can help accelerate
an EBM both in training and in inference in a number of different ways. For
this purpose, we choose a special case of Boltzmann machines called the
Restricted Boltzmann Machine (RBM) as it is a widely-used algorithm that is
heavily optimized for von Neumann architectures. RBMs (and its multi-layer 
variants) have found applications in specialized learning~\cite{salakhutdinov2009deep,nips2010_danl_george,
HJELM2014245,wang2014vehicle,iip2016_dingshifei,zhang2016deep,
abdel2016breast,al2021covid} and unsupervised learning~\cite{salakhutdinov2007restricted,larochelle2009exploring,erhan2010does,hinton2011discovering,fiore2013network,srivastava2013modeling,gouveia2017systematic,xie2018end,hernandez2018learning,fatemi2019joint,wang2020novel}.

\begin{figure}[ht]
\centering
\includegraphics[width=0.40\textwidth]{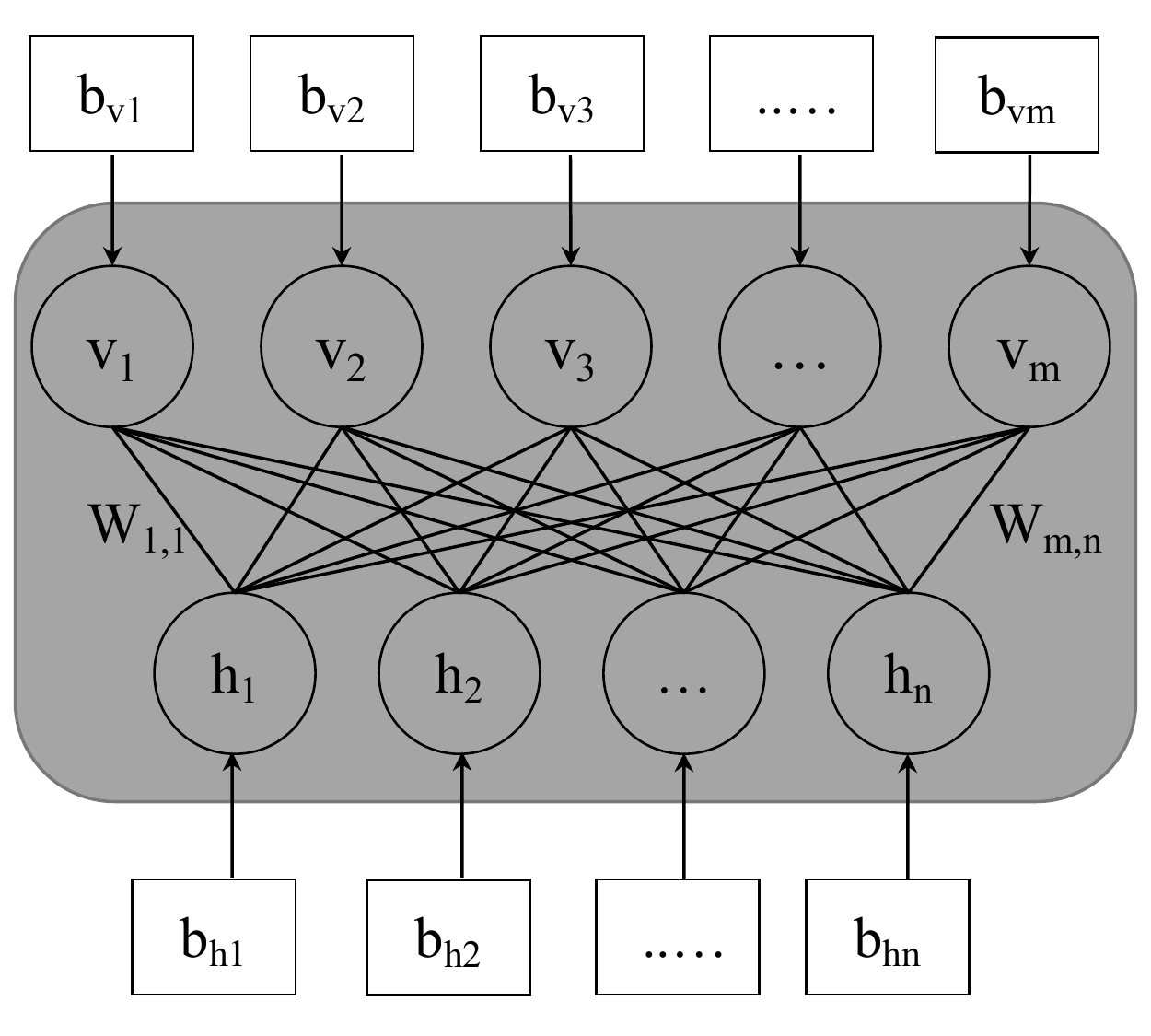}
\caption{Restricted Boltzmann machine (RBM)}\label{fig:rbm}
\end{figure}

An RBM has only connections between a visible node and a hidden 
node and no connections between two visible nodes or between two hidden nodes as shown 
in Figure~\ref{fig:rbm}. As a result, the energy function is as follows: \begin{equation}
\label{eqn:energy of RBM}
E(v,h) = -\sum_i^m \sum_j^n v_i W_{ij} h_j -  \sum_i^m b_{v_i} v_i -\sum_j^n b_{h_j} h_j
\end{equation} where $W_{i,j}$ is the coupling weight between visible unit $v_i$ and hidden 
unit $h_j$; and  $b_{v_i}$ and  $b_{h_j}$ are the biases for the 
corresponding visible  and hidden units.

Similar to other neural networks, RBMs can be stacked into a multi-layer 
configuration to form a deep network. Specifically, two common variants
are Deep Belief Networks (DBN) and Deep Boltzmann Machines (DBM)~\cite{salakhutdinov2009deep,hua2015deep}.
There are subtle differences between these variants and the simpler RBM.
For the sake of clarity, we will focus on discussing RBM and follow conventional
approaches when stacking multiple layers together. In other words, 
DBN/DBM-specific optimizations are outside the scope of this paper.

%% file: 3arch.tex
\section{Architecture}
\label{sec:arch}

In this section, we discuss the architecture design issues. We start with
our Ising machine substrate that will be the foundation for additional
architecture support (Sec.~\ref{ssec:brim}). Next, we discuss a design
where our modified Ising machine serves to provide acceleration for key
computation in an otherwise conventional approach to the training of
RBMs (Sec.~\ref{ssec:gibbs}). Finally we show that if we start from first
principles, we do not need to follow exactly the conventional
training algorithm and can better leverage the strength of our Ising substrate as a
practical Boltzmann sampler (Sec~\ref{ssec:boltz}). The result is more
flexibility in system design that ultimately leads to a more efficient
design.

\subsection{Ising machine substrate}
\label{ssec:brim}
\begin{figure}[htb]
\centering
    \includegraphics[width=0.42\textwidth]{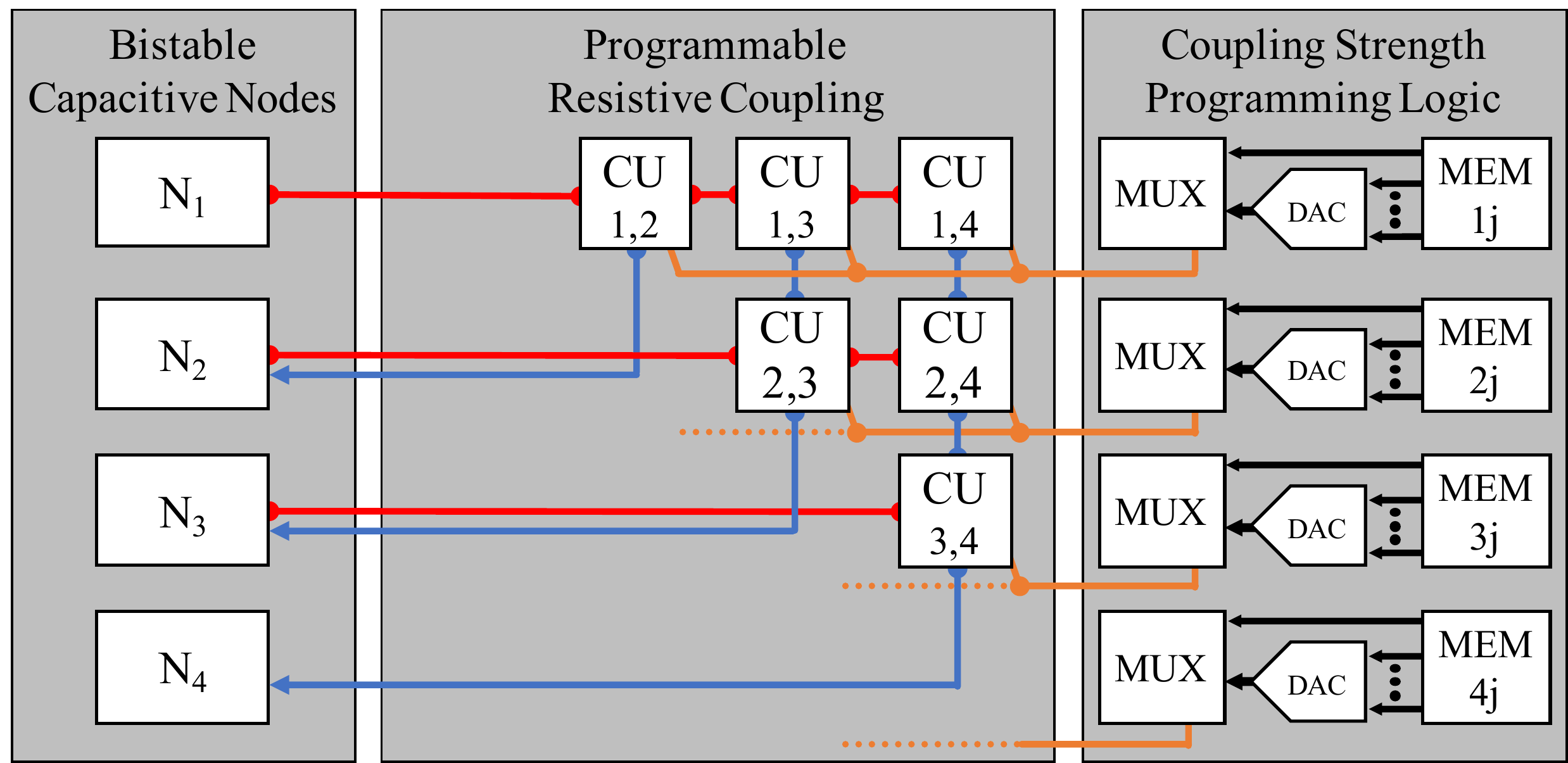}
    \caption{ High-level BRIM showing bistable capacitative nodes with
    programmable resistive coupling, and its programming logic. Note that
    between every pair of nodes (say, $N_1$ and $N_2$), we only show one
    bi-directional coupling units ($CU_{1,2}$), resulting in an upper triangular
    coupling network. In an equivalent implementation, the coupling unit
    may consist of two uni-directional parts, forming a symmetric layout.
    \label{fig:brim}}
\end{figure}

As we discussed earlier, a number of physical substrates can leverage nature
to perform optimization of an Ising formula. In principle, any such substrate
can be used for our purpose of accelerating a machine learning algorithm.
In practice, we aim for an integrated electronic design. This makes the
BRIM design~\cite{afoakwa.hpca21} a much more convenient baseline to use.
Figure~\ref{fig:brim} shows a high-level diagram of its architecture.

%\begin{figure*}[htb]\centering
%\includegraphics[width=0.6\textwidth]{rbrim_highlevel.pdf}
%\caption{Highlevel RBM showing visible and hidden nodes, with 
%clamping units to drive node biases, coupling mesh, and 
%programming utility.}
%\label{fig:rbrim}
%\end{figure*}

In this machine, each node consists of a capacitor and a feedback unit to
make the capacitor bistable ($1$ or $-1$). A mesh of all-to-all programmable 
resistors serve to express the Ising formula the system is trying to optimize.
When treated as a dynamical system, a Lyapunov analysis can be applied to the
differential equations governing the nodal voltages. It can be shown that
local minima of the Ising energy landscape are all stable states of the 
system~\cite{afoakwa.hpca21}. Put more simply, starting from a given
initial condition, this system of coupled nodes will seek a local minimum
without explicit guidance from a controller. Extra annealing control is 
needed to inject random ``spin flips" to escape a local minimum. This
is analogous to accepting a state of worse energy with non-zero probability
in simulated annealing~\cite{Kirkpatrick671}.

\begin{figure}[htb]
\centering
\includegraphics[width=0.49\textwidth]{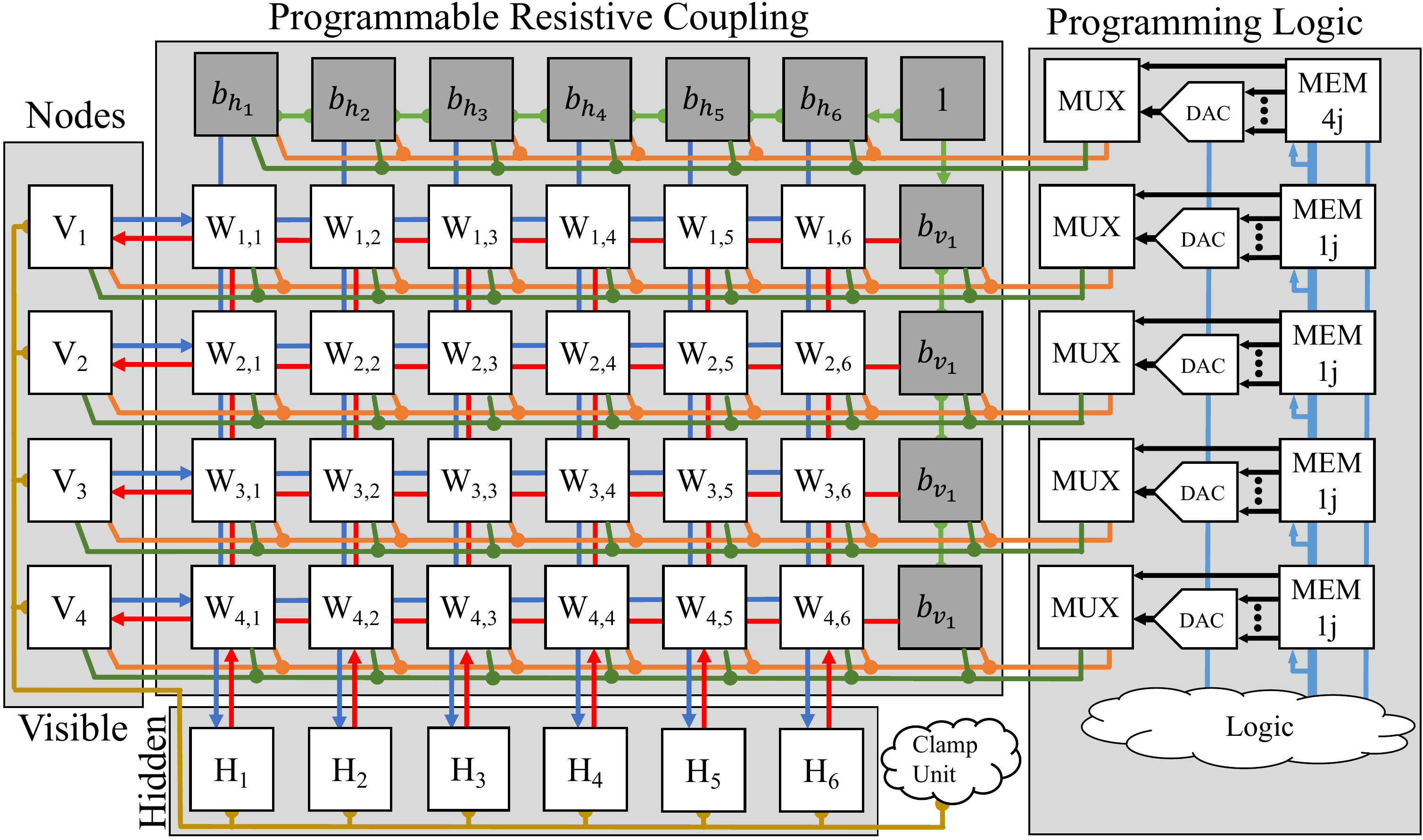}
\caption{High-level RBM showing visible and hidden nodes, with 
clamping units to drive node biases, coupling mesh, and 
programming logic.}
\label{fig:rbrim}
\end{figure}

In an RBM, the nodes are separated into a bipartite graph. For such a special
graph, the architecture for our Ising machine substrate can be slightly modified
to have nodes on two edges of the coupling network. In this structure, a visible 
node ($v_i$) can only be coupled to a hidden node ($h_j$)
as shown in Figure~\ref{fig:rbrim}. This layout significantly improves 
space efficiency compared to one that allows connection between all nodes.
As a concrete example, one of our benchmarks uses 784 visible nodes 
and 200 hidden nodes. Mapping them on a generic
all-to-all Ising substrate would need about 6 times more coupling units 
$((784+200)^2$ vs $784\times200)$.

Finally, the nodes of the Ising machine can be augmented in the following ways
to support operation of the RBM algorithm. In both training and inference, it
is common to clamp the visible nodes or the hidden nodes to certain value.
This can be easily achieved with a buffer connected to the nodal capacitor.
One important detail is that the inputs to the visible nodes are usually multi-bit
values and thus requires digital-to-analog conversion. Note that in
the baseline Ising machine, the vast majority of the area is devoted to the
coupling units as it scales with $N^2$ ($N$ being the number nodes). Thus most
addition to the node structure has a small impact to the system's complexity
and chip area.

% \todo{RKA here}
\subsection{Gibbs sampler architecture}\label{ssec:gibbs}

We first discuss a variant of RBM accelerator that is more traditional: we
are simply leveraging the Ising substrate to accelerate a portion of the
software algorithm that naturally suits the hardware. For convenience, we
name the accelerator design \emph{Gibbs sampler} (GS) as it follows (a variant 
of) the traditional Gibbs sampling-based algorithm~\cite{gelfand2000gibbs}. 
The overall algorithm is shown as Algorithm~\ref{lst:pseudocode}. 

\paragraph{The algorithm:}

Here we see that in the training loop (lines~\ref{lst:loop2_beg} to~\ref{lst:loop2_end}),
there are repeated calculations of $\langle v_{pos}^\top h_{pos} \rangle$ and 
$\langle v_{neg}^\top h_{neg} \rangle$. In a nutshell, the algorithm is a stochastic 
gradient descent and the training loop is calculating the stochastic gradient for 
every weight $W_{ij}$. In the so-called positive phase (lines~\ref{lst:vpos},~\ref{lst:hpos}),
a training sample $v_{pos}$ is clamped to the visible nodes; and a corresponding 
sample for the hidden nodes $h_{pos}$ is generated based on the conditional 
probability formula:
\begin{equation}
\label{eqn:probability_hidden}
P(h_j=1|v) = \sigma(b_{h_j} +  \sum_i W_{ij} v_i)
\end{equation}
where $\sigma(x)$ is the logistic function $1/(1+e^{-x})$.

In the negative phase (lines~\ref{lst:vneg},~\ref{lst:hneg}), a $k$-step Gibbs sampling 
is performed. Keeping the current hidden node values, we probabilistically generate 
a set of new visible nodes, with probabilities as follows. 
\begin{equation}
\label{eqn:probability_visible}
P(v_i=1|h) = \sigma(b_{v_i} +  \sum_j W_{ij} h_j) 
\end{equation}
From there, the updated visible nodes project back to generate 
updated hidden nodes, forming one complete step of the Markov Chain
Monte Carlo (MCMC) algorithm. In principle, one such step in the 
MCMC algorithm would make a rather poor sampling. In practice, a small
number of $k$ steps (\eg 5) are chosen to balance the cost and the quality of
the sampling. Such a $k$-step contrastive divergence algorithm is often
referred to as CD-$k$.

%\scalebox{0.9}{
%\begin{minipage}{2.1in} 

\begin{algorithm}[htb]
%\begin{boxedminipage}[4in]
    \small
    \caption{\small Contrastive divergence algorithm for training}
    \label{lst:pseudocode}
    \hspace*{\algorithmicindent} \textbf{Input:} training\_data, learning\_rate ($\alpha$), $CD_k$ , batch\_size ($bs$)\\
    \hspace*{\algorithmicindent} \textbf{Output:} weights (W), visible bias ($b_v$), hidden bias ($b_h$)
    \begin{algorithmic}[1]
        \State $W \leftarrow random(M \times N)$
        \State $b_{v} \leftarrow zeros(1 \times M)$
        \State $b_{h} \leftarrow zeros(1 \times N)$
        \State $number\ of\ batches$ = $\frac{size\ of\ training data}{batch\_size}$
        \State
        \For {$s = 1,2,\,\ldots\,, steps$}
           \For { $b= 1,2, \,\ldots\, ,number\ of\ batches$  }
              \For { $i = 1,2, \,\ldots\, ,bs$ } \label{lst:loop2_beg}
                \State $v_{pos}[i] \leftarrow training\_data$[$(b-1)*bs + i$]\label{lst:vpos}
                \State $h_{pos}[i] \leftarrow rand() < \sigma((v_{pos}[i] \times W) + b_{h})$ \label{lst:hpos}
                \State $h_{neg}[i] \leftarrow h_{pos}[i]$
                \For{$k = 1,2,\,\ldots\,, CD_{k}$}
                \label{lst:loop3_beg}
                  \State $v_{neg}[i] \leftarrow rand()<\sigma((h_{neg}[i] \times W^\top) + b_{v})$\label{lst:vneg}
                  \State $h_{neg}[i] \leftarrow rand()<\sigma(((v_{neg}[i] \times W) + b_{h})$\label{lst:hneg}
                \EndFor \label{lst:loop3_end}
              \EndFor \label{lst:loop2_end}
               \State $W \leftarrow W  + \alpha*(\langle v_{pos}^\top h_{pos}\rangle  - \langle  v_{neg}^\top h_{neg}\rangle )$
               \State $b_{v} \leftarrow b_{v} + \alpha * \langle v_{pos} - v_{neg}\rangle$
               \State $b_{h} \leftarrow b_{h} + \alpha * \langle h_{pos} -h_{neg} \rangle$
            \EndFor  
            %\State $\alpha_t \leftarrow decrease(\alpha_t)$
        \EndFor
    \end{algorithmic}
%\end{boxedminipage}
\end{algorithm}

\paragraph{Architectural support:}

At every learning step, the current weight matrix $[W_{ij}]_{M\times N}$ 
will be programmed to the coupling array such that the resistance at each 
unit $R_{ij}$ is proportional to $\frac{1}{W_{ij}}$. This step is just like
programming the optimization formula in a standalone Ising machine. If we
further clamp one set of nodes (say, visible) to fixed values, 
each coupling unit will produce a current equal to the voltage of the visible 
node divided by the resistance of the programmable coupling unit, which is 
equivalent to multiplying the corresponding weight in the matrix. Each hidden 
node, therefore, sees the sum of the current in the entire column. Used this 
way, the coupling array is effectively producing a vector-matrix multiplication
operation (in the analog domain).

At this stage, rather than reading out the resulting currents, we opt to feed
the current through a non-linear circuit that produces the effect of the
logistic function. In fact, a modified inverter can
approximate the function admirably. Finally, the
output of the logistic function is the probability of node being 1. This can
also be supported with a relatively straightforward circuit: a comparator with
the other input being fed with pseudo random voltage level. 
Details of circuit design and issues can be found in Appendix B.

\paragraph{Operation:}

With the architectural support described, much of the training loop in 
Algorithm 1 will be offloaded to the hardware. The remaining operations
will be carried out on digital functional units such as a TPU.
The overall operation sequence is as follows:
\begin{enumerate}[nolistsep]
    \item Initialization on TPU; 
    \item Programming the (initially random) coupling matrix and biases to the Ising substrate.
    \item Clamping the visible units to a training sample ($v_{pos})$;
    \item Read out the hidden units ($h_{pos}$) after the Ising substrate 
    finishes operation;
    \item Perform the equivalent of $k$-step Gibbs sampling by initializing the 
    hidden units and allowing the Ising substrate to evolve and produce state samples.
    \item Read out the final value from the visible ($v_{neg}$) and hidden units ($h_{neg}$).
    \item Repeat steps 3 to 6 to collect the samples for calculating gradient descent. %\st{calculate gradient based on minibatch average$(\langle v_{pos}h_{pos}\rangle  - \langle  v_{neg}h_{neg}\rangle )$}
    \item Compute (on TPU) the new coupling matrix and biases by calculating gradient using the samples collected according to the following formula $(\langle v_{pos}^\top h_{pos}\rangle  - \langle  v_{neg}^\top h_{neg}\rangle )$.
    \item Repeat from step 2 for subsequent minibatches. %\st{learning steps}.
\end{enumerate}

\subsection{Boltzmann gradient follower architecture}\label{ssec:boltz}

The system described in Sec.~\ref{ssec:gibbs} represent an improvement over
digital units as we shall see later. But this is largely due to the efficiency
gain from approximate analog implementations. The benefit of nature-based
computing is often much greater when we avoid frequent interactions between
specialized hardware and a more general-purpose computing substrate.
For this to happen, we need to have a deeper understanding
of the \emph{intention} of the algorithm.

\paragraph{The goal of the MCMC algorithm:}
With RBM, the goal is to capture the training data with a probability distribution
model \ie of all possible inputs, those in the training set should have high probabilities. 
The probability of an input is exponentially related to the energy
of that input as in Boltzmann distribution (hence the name):
\begin{equation}
\label{eqn:boltz_prop}
P(v,h) \propto e^{-E(v,h)} \\
\end{equation}
Clearly the probabilities need to sum up to one, thus the equation for 
the probability of a particular state $(v,h)$ is:
\begin{equation}
P(v,h) =  \frac{1}{Z} e^{-E(v,h)};\ \   Z=\sum_{v,h} e^{-E(v, h)}
\label{eqn:boltz_part}
\end{equation}
When we say the machine captures the training data, we mean the machine's
model (weights and biases) maximizes the probability of all $T$ training 
samples. In other words, it maximizes $\prod_{t=1}^{T} P(v^{(t)})$ where
$v^{(t)}$ is the $t^{th}$ training sample, or equivalently the sum of the log
probability: $\sum_{t=1}^T log(P(v^{(t)}))$. Note that $P(v)=\sum_h P(v,h)$.
Since the probability is a function
of the parameters (coupling weights and biases), we follow the gradient of
each parameter. For the coupling parameter, the gradient is as follows:
\begin{equation}\label{eqn:loglikelihood}
\small
\pdv{\sum_{t=1}^TlogP(v^{(t)})}{W_{ij}} = \sum_{t=1}^T\pdv{[log(\sum_h e^{-E(v^{(t)},\ h)})-log(Z)]}{W_{ij}}
\end{equation}

For notional clarity, we now look only at the contribution of one training sample ($u=v^{(t)}$)
to the gradient and focus on the first part in the numerator of RHS of Eq.~\ref{eqn:loglikelihood}:
\begin{equation}
\footnotesize
\begin{split}
\pdv{log(\sum_he^{-E(u, h)})}{W_{ij}} &= \frac{1}{\sum_he^{-E(u,h)}}\pdv{\sum_he^{-E(u,h)}}{W_{ij}} \\
&= \frac{1}{\sum_he^{-E(u,h)}}\sum_he^{-E(u,h)}\pdv{(-E(u, h))}{W_{ij}} \\
&= \frac{ \sum_he^{-E(u,h)}u_ih_j}{\sum_he^{-E(u,h)}} =\langle u_ih_j\rangle_{data}
\end{split}
\end{equation}

Here the notation $\langle\cdot \rangle_{data}$ means the expectation 
with respect to the data, \ie keeping the data constant ($u$) and averaging over
all possible $h$.

If we follow the same steps, the second part of the gradient ($\pdv{log(Z)}{W_{ij}}$) gives us:
\begin{equation}
% \pdv{log(Z)}{W_{ij}} = \frac{ \sum_{h,v}e^{-E(v,h)}v_ih_j}{\sum_{h,v}e^{-E(v,h)}} =\langle v_ih_j\rangle _{model}
\pdv{log(Z)}{W_{ij}} = \frac{ \sum_{v,h}e^{-E(v,h)}v_ih_j}{\sum_{v,h}e^{-E(v,h)}} =\langle v_ih_j\rangle _{model}
\end{equation}
Here the notation $\langle\cdot \rangle_{model}$ means the expectation with respect 
to the entire state space given by the current model (coupling parameters and biases).

%\begin{figure}[htb] %{r}{1.35in}
%    \includegraphics[width=0.40\textwidth]{brim_energy_cdf.pdf}
%    \caption{Distribution of BRIM energy state}
%    \label{fig:ising_vs_Brim_dist}
%\end{figure}

As we can see, to calculate \emph{any} parameter's gradient, we need to
calculate the expectation of a large number of states, which is impractical.
The common solution is an MCMC algorithm (\eg CD-$k$), just like simulated
annealing in solving an Ising formulation problem. Both algorithms are
inspired by Boltzmann statistics. The Ising machine substrate that we
use \emph{directly embodies} such statistics and can thus perform sampling of
the state space with extraordinary efficiency. To see the efficiency, let us
look at a concrete example. For a 784$\times$200 RBM, the $k$-iteration software Gibbs sampling
(lines~\ref{lst:loop3_beg} to \ref{lst:loop3_end} in Algorithm 1) is achieved by taking $s=k*(784+200)$ steps on the Markov chain and takes roughly
$k\times10^6$ instructions. This is equivalent 
to going through a trajectory on our dynamical system with roughly $s$ phase points,
each taking roughly a dozen picoseconds on average.

%the hardware on the orders of 2ns.}
%\note{try to use sampling example}

%\red{ In a 1000-node Ising formula optimization problems,
%the state-of-the-art simulated annealing algorithm~\cite{Isakov} takes $10^5$
%instructions to mimic one step in the random walk. For an Ising machine like
%BRIM, this is equivalent to a spin flip that occurs roughly every 10ps. Similarly
%for a 784$\times$200 RBM, the $k$-step software Gibbs sampling
%(lines~\ref{lst:loop3_beg} to \ref{lst:loop3_end} in Algorithm 1) takes roughly
%$k\times10^6$ instructions, whereas a similar quality sampling will take
%the hardware on the orders of 2ns.}
%\note{try to use sampling example}

%\footnote{ Indeed, in both cases, the Markov chains are time-inhomogeneous. There is a subtle difference between the chain's time-inhomogeneity. In simulated annealing of an Ising formula, the coupling parameters do not change while the temperature gradually reduces. Whereas in contrastive divergence, the effective temperature remains the same while the coupling parameters change.} 

%We leave the theoretical analysis as future work and just show an empirical observation here. Figure~\ref{fig:ising_vs_Brim_dist} shows the cumulative distribution of the  energy of states visited by our Ising substrate and a fitted curve of a Boltzmann distribution for the same set of energies. The agreement suggests that this Ising substrate can be considered as a Boltzmann sampler. 

\paragraph{Architectural support:}
When we initialize the Ising substrate to some initial condition, it will proceed
to traverse through the energy landscape directed by both the system's governing
differential equations and the annealing control. This has the effect of ``sampling"
the state space and arguably produces samples better than the algorithmic random 
walk in CD-$k$. However, there is a problem: the production of the samples is 
much faster than the host computer can access and postprocess them to obtain the 
expectations. We propose, therefore, a more direct approach, where the
sampled expectations ($\langle v_ih_j\rangle_{data}$ or $\langle v_ih_j\rangle_{model}$)
are directly added to or subtracted from the model parameter (\eg $W_{ij}$) inside
the Ising substrate, without involving the host.

In the CD-$k$ algorithm, learning is as follows, where
$\langle\cdot\rangle_{s^\pm}$ indicates the expectation over a set of positive 
($s^+$) or negative ($s^-$) phase samples:
\begin{equation}
\label{eqn:weight_adust}
W_{ij}^{t+1} = W_{ij}^{t} +\alpha\Big(\langle v_{i}h_{j}\rangle_{s^+}-\langle v_{i}h_{j}\rangle_{s^-}\Big)
\end{equation}
Here the expectations are accumulated over a minibatch of (say, $n=500$) 
samples before being used to update the parameters to the next value. The
choice of $n$ is usually a matter of convenience for the software.
For us, the minibatch size is effectively one ($n=1$),
which requires a correspondingly smaller $\alpha$ (roughly 500$\times$ 
less than that needed for $n=500$) for a similar effective learning
rate.
In this setup, the hardware will then take one positive phase sample and
increment $W_{ij}$ based on this sample, then take a negative phase sample and
decrement $W_{ij}$ based on that sample. 
This is ideal for hardware implementation as no expensive storage is needed
to accumulate statistics for calculating expectations.
In our baseline Ising machine, each
$W_{ij}$ is physically expressed by the conductance of a configurable
resistor, which is implemented by a transistor with variable gate-source
voltages. Increasing and decreasing $W_{ij}$ can thus be achieved by raising
or lowering the gate voltage using a charge-redistribution circuit
shown in Fig.~\ref{fig:boltz_samp} with more details in Appendix~\ref{sec:circuit}.

A fine point can be made here as to whether 
our learning algorithm produces a biased estimator or not. 
Our empirical analysis shows 
estimation bias appears to have no effect on the ultimate accuracy measures.
If anything, our modifications seem to reduce bias from the commonly used
algorithms as detailed in Appendix A.

For negative phase samples, we use $p$ different initial conditions for the
hidden units ($h^{(l)}, l=1..p)$ to allow $p$ independent random walks. These
are often referred to as $p$ particles. After each positive phase, we load
one of the particles (say, $h^{(3)}$); perform annealing of the Ising machine
(equivalent to the random walk of the von Neumann algorithm); take
a sample of $v_{i}h_{j}$ (negative phase); and store the resulting hidden 
unit values back to $h^{(3)}$ for persistence~\cite{tieleman2008training}.

The parameters are physically expressed by the conductance of configurable
resistors. The resistors are implemented by transistors with variable gate-source
voltages. Increasing and decreasing the parameters can be achieved by raising
or lowering the gate voltage. This in turn is done by a charge-redistribution circuit
detailed in Appendix B.

The resulting operation can be thus characterized by the following equations:
\begin{equation}
\begin{split}
W_{ij}^{t+\frac{1}{2}} &= W_{ij}^{t} +f_{ij}\Big(\alpha\langle v_{i}h_{j}\rangle_{s^+}
\Big)\\
W_{ij}^{t+1} &= W_{ij}^{t+\frac{1}{2}} -f_{ij}\Big(\alpha\langle v_{i}h_{j}\rangle_{s^-}
\Big)
\end{split}
\label{eqn:bgf}
\end{equation}
Constrasting Eq.~\ref{eqn:bgf} with Eq.~\ref{eqn:weight_adust}, we
can summarize the effective changes to the original algorithm as follows: \begin{enumerate}
    \item our parameters are updated mid-step: positive phase samples ($s^+$) are
    taken under ($W^t$), resulting in updated parameters $W^{t+\frac{1}{2}}$, under
    which negative phase samples ($s^-$) are taken. In the original algorithm, both
    samples are taken under the same parameter ($W^t$).
    \item the increments are achieved by hardware adjustment, which has both non-linearity and variations. This effect is captured by the function $f_{ij}(\cdot)$.
    \item for convenience of the hardware, the effective mini-batch size is 1.
    \end{enumerate}
    
Finally, since the entire learning is now conducted inside the (augmented) Ising 
substrate, the coupling unit is larger (Fig.~\ref{fig:boltz_samp}). Furthermore, the 
trained results need to be read out at the end of the learning process, this
requires extra analog-to-digital converters (ADC) which are expensive
converters. Nevertheless, keep in mind that they are only used once at the end of 
the entire algorithm.

In short, though the architecture needs some non-trivial new circuits and small
modifications to the operating algorithm, it carries out the \emph{intention}
of a traditional software implementation. As we will see, the resulting quality
is no different from a software implementation even under non-trivial noise
considerations.

\begin{figure}[htb]\centering
\includegraphics[width=0.45\textwidth]{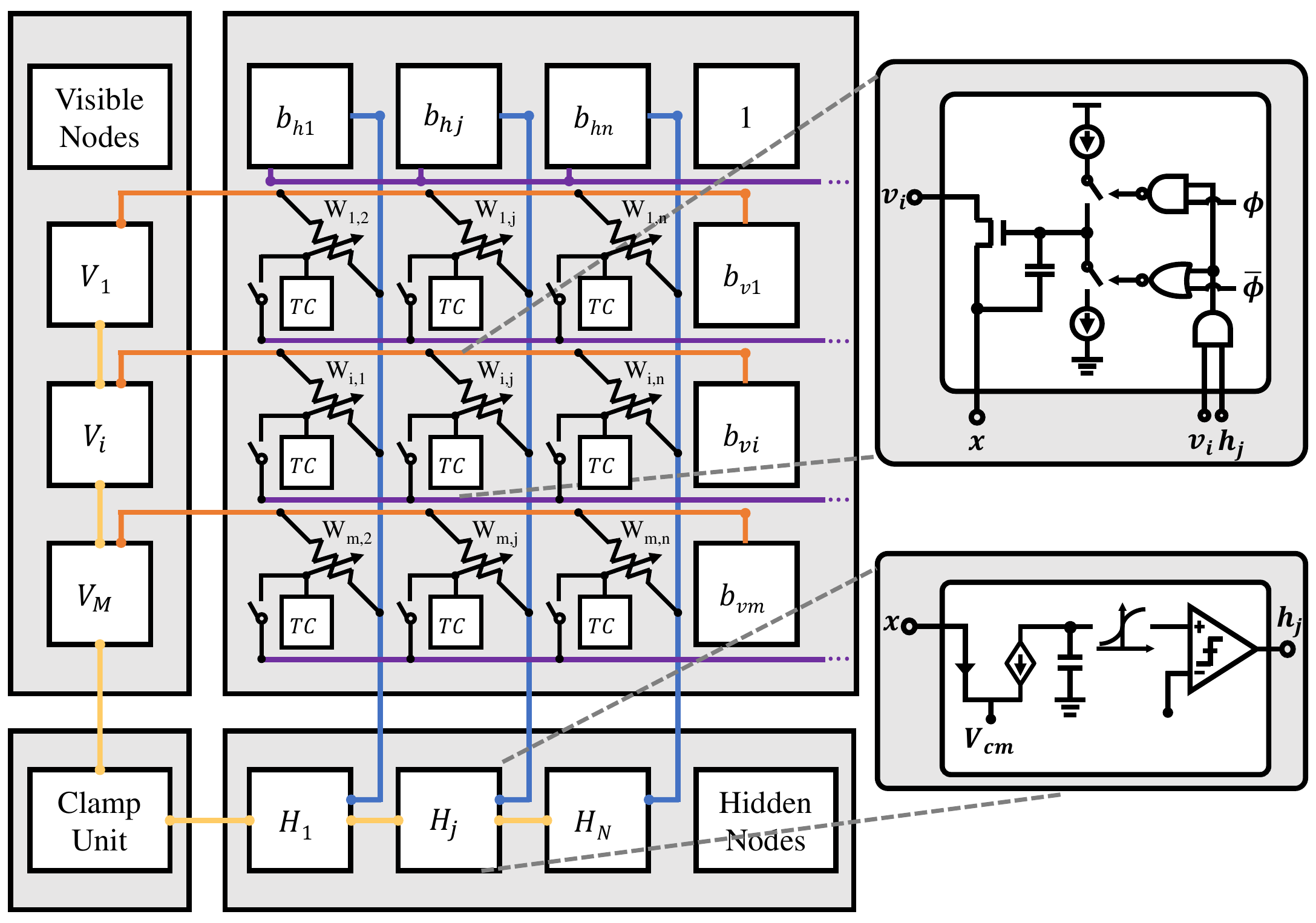}
    \caption{An architecture diagram of Boltzmann gradient follower. }
    \label{fig:boltz_samp}
\end{figure}

\paragraph{Operation:}

Figure~\ref{fig:boltz_samp} shows the proposed Boltzmann gradient follower
architecture. The key addition is to the coupling unit where the programmable
resistor serving as the weight can be adjusted in place. 
With this architecture, the digital (host) computer takes 
a much more peripheral role, mostly setting the system up, feeding 
data at a fixed frequency, and finally reading the results. The operation 
can be described as follows:

\begin{enumerate}[nolistsep]
    \item Initialize the weights and biases.\footnote{Note that for common practice, the
    weights are initialized to small random values. This could certainly be
    implemented by the hardware itself. But programmable initial conditions may
    be useful for special purposes in training (\eg research) and certainly for inference.}
    \item The host would send training samples to latches at the visible units.
    \item The machine will clamp the data, wait for a predetermined time for the
    hidden units to settle. The resulting sample $\langle v_{i}h_{j}\rangle_{s^+}$ will 
    increment $W_{ij}$.
    \item The machine will then load one of $p$ particles and start annealing
    process.
    \item After annealing, the resulting sample $\langle v_{i}h_{j}\rangle_{s^-}$
    will decrement $W_{ij}$.
    \item The process (2-5) is repeated for a programmable number of learning steps.
    Then the ADCs will read out the coupler control voltages one column at a time.
\end{enumerate}

%% file: 5eval.tex
\section{Experimental Analysis}\label{sec:eval}

To better understand the characteristics of our augmented Ising substrate
for EBMs, we compare various metrics between our proposed design and
TPU~\cite{jouppi2017datacenter}. It is worth noting up front that an
evolutionary perspective is needed when viewing the results. Both digital
numerical architectures and nature-based computing systems will evolve. Design
refinement can bring significant changes to these metrics. More importantly,
the line between the two groups will continue to blur and cross pollination
is very much an intended consequence.

We next describe the experimental setup including benchmark RBMs and DBN
models; then show quantitative comparisons between a TPU
and a noiseless model of our analog design in terms of energy efficiency and
throughput; and finally present a few in-depth analyses to help understand the how
the system behaves.

\subsection{Experimental Setup}

\paragraph{Datasets and metric:}To evaluate the systems discussed, we trained them on different
applications like image classification, recommendation systems and
anomaly detection. For image classifications we use many datasets
including handwritten digits (MNIST)~\cite{deng2012mnist},
Japanese letters (KMNIST)~\cite{clanuwat2018deep}, fashion
images (FMNIST)~\cite{xiao2017fashion}, extended handwritten
alphabet images (EMNIST)~\cite{cohen2017emnist}, 
toy dataset (SmallNorb)~\cite{lecun2004learning}, and
medium color images (CIFAR10)~\cite{krizhevsky2009learning}. %and larger images of 101 categories of objects (Caltech 101)~\cite{kaggle02}.
Image sizes range from 28x28 gray scale (NIST), to 32X32 color image (CIFAR10).
RBM and Deep Belief Network (DBN) are directly attached to the images in the NIST
datasets while Convolution RBM
algorithm~\cite{coates2011analysis} is used for CIFAR10 and SmallNorb datasets.

To train RBM as a recommendation system~\cite{verma2017collaborative}
and anomaly detector \cite{Pumsirirat2018} we use 100k MovieLens
dataset~\cite{harper2015movielens} and "European Credit Card Fraud
Detection"~\cite{kaggle01} dataset respectively. The learning rate used to train these
models is 0.1 and size of RBM and DBN configurations are shown in
Table~\ref{tabel:rbm_sizes}. Since RBMs are unsupervised models, one way
to quantify the quality of training is the average log probability of the
training samples which can be measured using annealed importance sampling~\cite{salakhutdinov2008quantitative}. We also report some common metrics like
classification accuracy using logistic regression layer at the end for image
classification, mean absolute error (MAE) of test data and projected data for
recommendation systems and area under Receiver operating characteristic (ROC)
curve for Anomaly detection. 

\begin{table}[ht]\centering
\caption{Dataset parameters of different types of Neural Networks used in evaluation.}
\label{tabel:rbm_sizes}
\resizebox{\columnwidth}{!}{
\begin{tabular}{|c|c|c|} 
\hline
\bf{Datasets} &  \bf{RBM}    & \bf{DBN-DNN}~\cite{hinton2012deep}    \\  %& \bf{DBM} \\ 
\hline\hline
MNIST         &   784-200    &  784-500-500-10   \\ %&  784-500-500-10    \\ 
\hline
KMNIST        &   784-500    &  784-500-1000-10  \\ %&  784-500-1000-10   \\ 
\hline
FMNIST        &   784-784    &  784-784-1000-10  \\ %&  784-500-1000-10   \\
\hline
EMNIST        &   784-1024    &  784-784-784-26   \\ %&  784-784-784-27    \\
\hline
CIFAR10       &   108-1024    &  -   \\ %&  784-784-784-27    \\
\hline
SmallNorb     &   36-1024     &  -   \\ %&  784-784-784-27    \\
\hline
%Caltech101    & - & 9216-5000-10000 \\
%\hline
Recommendation systems  &   943-100     &  -   \\ %&  784-784-784-27    \\
\hline
Anomaly detection     &   28-10     &  -   \\ %&  784-784-784-27    \\
\hline
\end{tabular}
}
\end{table}

\paragraph{Modeling:}Behavioral models of Gibbs sampler (GS) and Boltzmann gradient follower (BGF) were developed using Matlab. All circuits are modeled in Cadence 45$nm$ using Generic Process Design Kit (GPDK045) to obtain power and area. We designed a 32x32 node Boltzmann gradient follower in cadence to test the behavioral models. We used these behavioral models to obtain performance and noise analysis. We assume the system has enough nodes to fit the largest problems in the set. Thus execution time is just the product of the number of iterations and the cycle count per iteration. Anything not carried out on our hardware is done on the host machine, which we take to be the same TPU as the baseline. The area and power consumed by the TPU unit were obtained from \cite{jouppi2017datacenter}, which uses a $28 nm$ technology. We used 8-bit precision ADCs to read out the final trained weights. To input training data, we used 8-bit precision DTCs. We adopt DTCs, and ADCs based on the measurement-validated designs in the following papers ~\cite{7765065},~\cite{6607254}, respectively.

\subsection{Execution Speed}

\begin{figure}[htb]\centering
\includegraphics[width=0.45\textwidth]{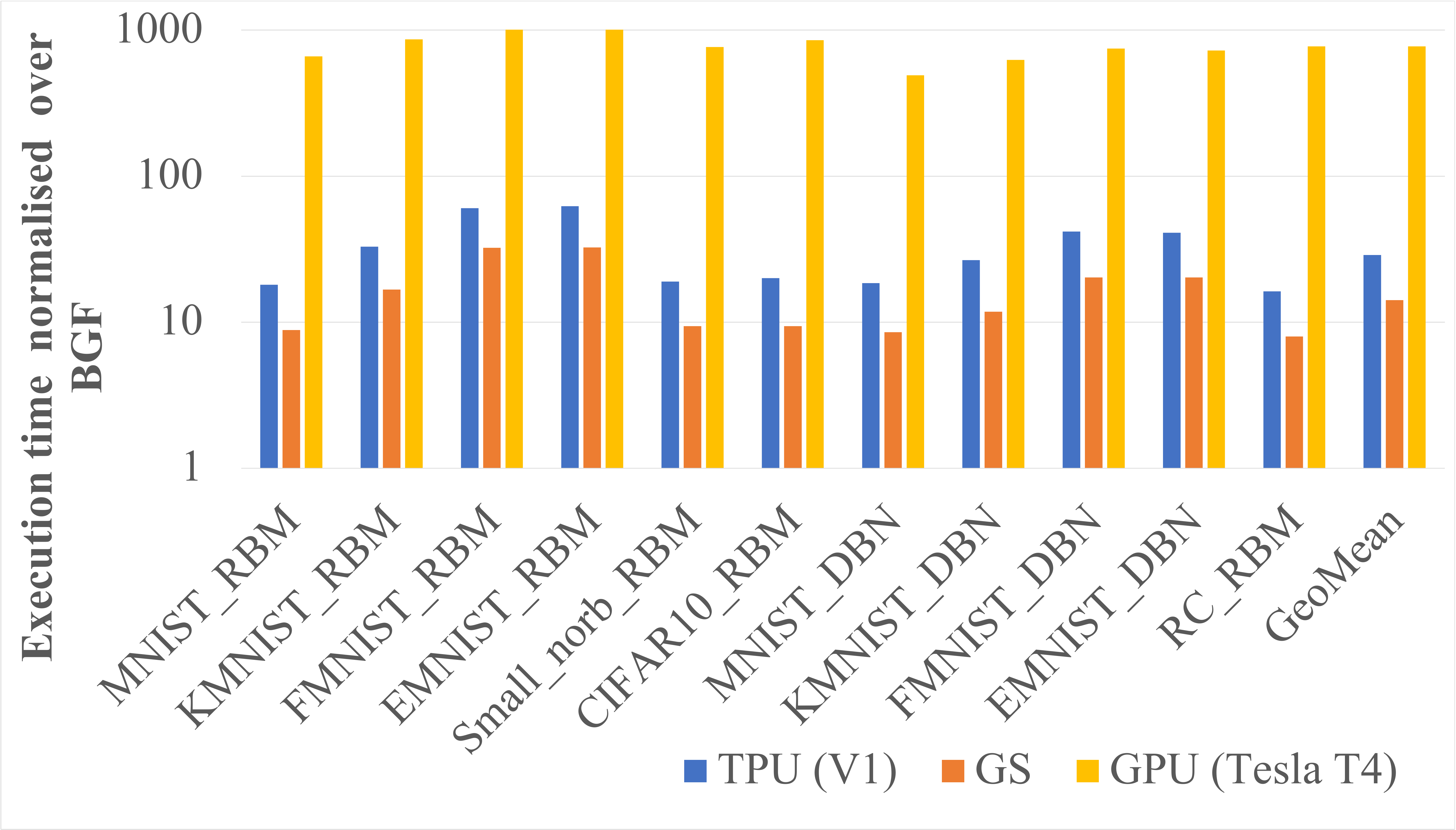}
\caption{Execution time normalized to that of BGF
for different RBMs and image batch size of 500.\label{fig:eval_perf}}
\end{figure}

We first consider the execution speed of the two architectures discussed: the
Gibbs sampler (GS) and the Boltzmann gradient follower (BGF). The operating 
frequency is 1GHz for digital portions for both. For a comparison, we include TPU
(v1) and a GPU. Fig.~\ref{fig:eval_perf} shows the execution time of different benchmarks 
all normalized to that of BGF. 
As we can see, BGF has 29x geometric mean speedup over TPU, whereas GS has 2x.
However, this is not meant as an apple-to-apple comparison and we need to contextualize 
the results:

First, TPU/GPU are meant to be more general-purpose and can handle
arbitrary sized workloads. In their current incarnation, BGF and GS 
have a fixed capacity and are limited in problem size they can process. 
That said, most datasets we used comfortably fit inside a small die. 
Other than
increasing die size for larger capacity, we note that scaling beyond
a single chip's capacity is feasible and part of the community's on-going 
research~\cite{sharma2022increasing}.

Second, communication between the Ising substrate and host is fully accounted
for and amounts to about a quarter of time GS spends waiting for host.
Indeed, the advantage of BGF over GS is precisely the removal of
the Amdahl's bottleneck of computing and communication of the host: 
BGF now automatically follows the gradient and only communicates end results
to the host. Nevertheless, the fundamental efficiency of the Ising substrate is
still an important factor as we show in Table~\ref{tabel:maxops}.

\subsection{Energy Consumption}

A primary source of fundamental efficiency of an Ising substrate comes from the fact that many algorithms
are \emph{mimicking} nature, whereas our hardware directly embodies that nature. For instance, in a typical step of MCMC, flipping one node requires roughly $O(N)$
multiply-accumulate (MAC) operations followed by some probability sampling. 
Ignoring the probability sampling, which may be much more costly, 
just a MAC operation costs on the order of a pJ.
For the problems discussed, $N\approx 1000$. So one such flip requires the order
of nJ using conventional digital computation. In contrast, in BRIM, flipping a
node involves a distributed network of currents charging/discharging a nodal
capacitor. With nodal capacitors on the order of 50 fF and 
a voltage of roughly 1V, flipping of a node takes on the order of 100 fJ. 
Thus, an Ising substrate has the potential to be about 4 orders of magnitude more efficient compared to a conventional computational substrate.

\begin{figure}[htb]\centering
\includegraphics[width=0.45\textwidth]{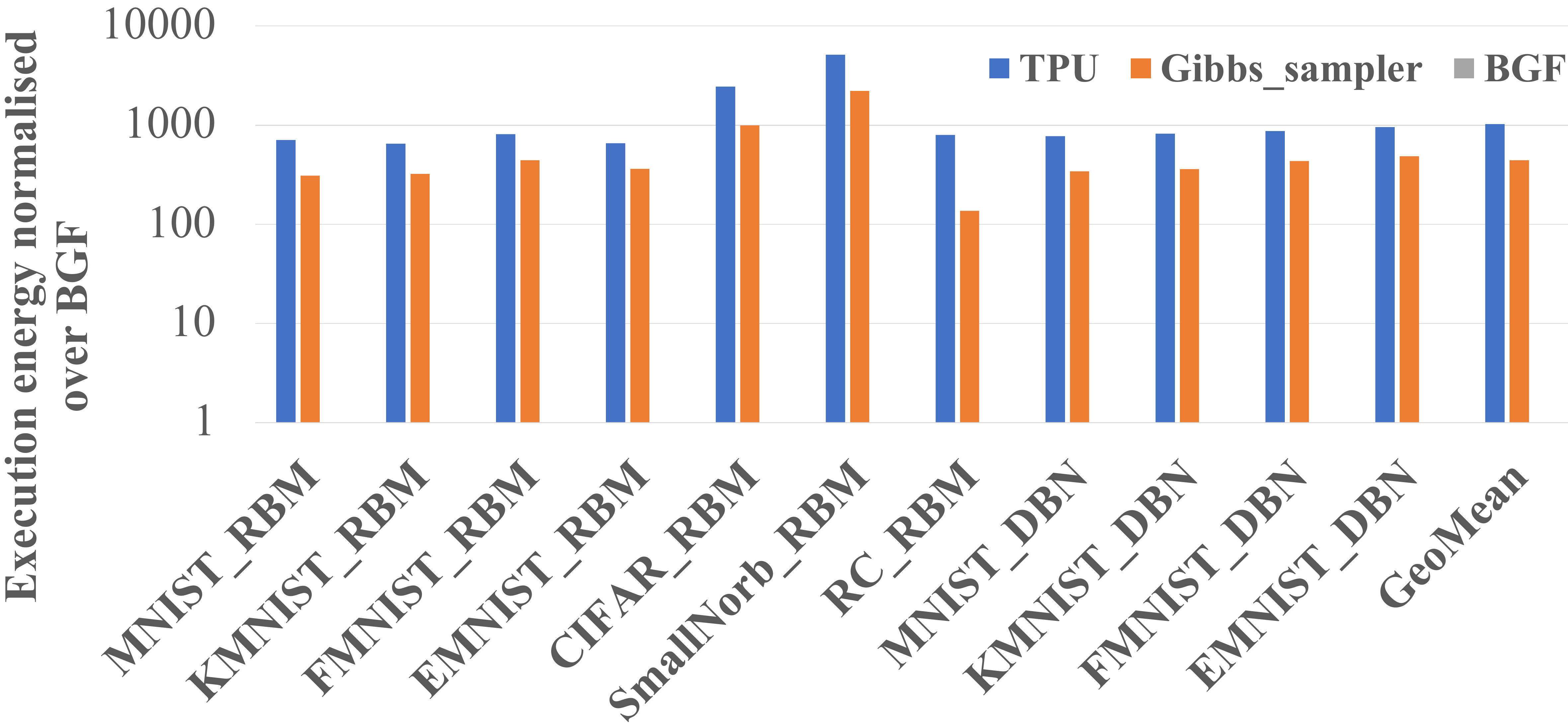}
\caption{Energy consumption of TPU and GS over various benchmarks 
normalized over BGF for image batch size of 500.\label{fig:eval_energy}}
\end{figure}

Regarding the entire system, the energy savings will depend on many
factors not included in the simplified analysis above. With the technology difference in mind, we can compare the energy consumption of different benchmarks shown in Figure~\ref{fig:eval_energy}. All energy results are normalized to that of the Boltzmann gradient follower.

Our accelerators are more energy efficient in the effective operations they carry
out than digital TPU operations prescribed by the algorithm. Overall, we
see improvements of around 1000x for BGF. Two qualitative points are worth
mentioning. On the one hand, digital circuit can still improve and more
efficient designs can further reduce per-operation cost. On the other hand,
the type of short random walks performed on our Boltzmann gradient follower
architecture are not the most efficient use of the Ising substrate. Further
algorithm innovation may well find applications that fully utilize the ability
of ours and similar nature-based systems.

\begin{table}[htb]
 \centering
 \caption{Area and power of sub-units of Gibbs sampler and Boltzmann gradient follower at different number of nodes(N). CU = Coupling unit; SU = Sigmoid unit; RNG = Random number Genrator; DTC= Digital to Time Converter  }
 \label{tab:building_blks}
 \resizebox{\columnwidth}{!}{
 \setlength{\tabcolsep}{4pt}
 \begin{tabular}{|c|c|c|c|c|c|c|}
 \hline
 \multirow{2}{*}{\thead{Components / \\Nodes}} & \multicolumn{2}{c|}{$400\times 400$} & %
     \multicolumn{2}{c|}{ $800\times 800$ } & \multicolumn{2}{c|}{$1600\times 1600$}\\
 \cline{2-7}
  & \thead{area \\ $(mm^2)$} & \thead{power \\ (mW)} & \thead{area \\ $(mm^2)$} & \thead{power \\ (mW)} &  \thead{area \\ $(mm^2)$}  & \thead{power \\ (mW)}  \\
 \hline
  CU (Gibbs) $(N^2)$ & 0.03   & 30    & 0.12     & 120    & 0.48    & 480 \\
  CU (BGF) $(N^2)$   &  1.28  & 36    & 5.12     & 144    & 20.5    & 576\\
  \hline
  SU $(N)$           & 0.0024 & 3.26  & 0.0048    & 6.52  & 0.0096  & 13\\
  \hline
 Comparator $(N)$    & 0.024  &  2    & 0.048     & 4     & 0.96    & 8\\
 \hline 
 DTC $(N)$           & 0.0004 & 7     & 0.0008    & 14    & 0.0016  & 28\\
 \hline  
 RNG $(N)$           & 0.007  & 18.24 & 0.014   & 36.48   & 0.028   & 72.96\\
 \hline
 Total(Gibbs)        &  0.065 & 60.5  & 0.19    & 181     & 1.5     & 601.96 \\
 Total(BGF)          &  1.32  & 66.5  & 5.19    & 205     & 21.5    & 700 \\
 \hline

 \end{tabular}
 }
 \end{table}

Finally, we look at the chip area estimates. Again, with the technology
difference this is also not a direct comparison. Nevertheless, our baseline
28nm TPU takes about 330$mm^2$ with 24\% of it being the MAC array. In
comparison, the area and power of BGF and its building blocks are shown
in Table~\ref{tab:building_blks}. Because the coupling unit is by far the
most numerous elements (roughly $O(N^2)$ vs $O(N)$ for other units), the
area is largely determined by the coupling units. Assuming a $1600\times1600$
array, a Boltzmann gradient follower costs about 21$mm^2$ which is small when compared to TPU area. %and Boltzmann gradient follower, the area is about 2$mm^2$, making it about 4.8\% of total host TPU area. 
As
shown in Table~\ref{tabel:maxops}, BGF are much more efficient -- in this
specialized algorithm -- than TPU
or the state-of-the-art computational accelerators. 

\begin{comment}
\begin{table} [ht]
\centering
\caption{Comparison between different accelerators. }
\vskip -10pt
\label{tabel:maxops}
\resizebox{0.8\columnwidth}{!}{
\small
\begin{tabular}{|c|c|c|} 
\hline
\bf{Accelerators} &  \bf{$TOPS/mm^2$} &  \bf{$TOPS/W$} \\ 
\hline\hline
 $TPU(v\_1,v\_4)$ ~\cite{jouppi2017datacenter,jouppi2023tpu}   &  1.16, 1.91  & 2.30, 1.62  \\ 
\hline
\hline
$TIMELY$ ~\cite{li2020timely}  &  38.3 & 21.0  \\   
\hline
$BGF(1600 \times 1600)$     & 243 & 8490 \\ 
\hline
\end{tabular}}
\end{table}
\end{comment}

\begin{table} [ht]
\centering
\caption{Comparison between different accelerators. }

\label{tabel:maxops}
\resizebox{0.8\columnwidth}{!}{
\small
\begin{tabular}{|c|c|c|} 
\hline
\bf{Accelerators} &  \bf{$TOPS/mm^2$} &  \bf{$TOPS/W$} \\ 
\hline\hline
 $TPU(v\_1,v\_4)$ ~\cite{jouppi2017datacenter,jouppi2023tpu}   &  1.16, 1.91  & 2.30, 1.62  \\ 
\hline

%$TPU(v4)$ ~\cite{jouppi2023tpu}   & 1.91   & 1.62  %\\ 
\hline

$TIMELY$ ~\cite{li2020timely}  &  38.3 & 21.0  \\   
\hline
$BGF(1600 \times 1600)$     & 119 & 3657 \\ 
\hline
\end{tabular}
}
\end{table}

\subsection{Impact of algorithmic change}
\label{ssec:impact_of_algo}

From a first-principle analysis, our Boltzmann gradient follower architecture
simply implements a different style of stochastic gradient descent. It will
not give us the exact same trained weights, but should provide similar
solution quality to the end user. We now analyze numerically the change
resulted from the algorithmic change. We use two metrics as discussed before:
the average log probability of the training samples and classification
accuracy. 
\begin{figure}[htb]
  \centering
 \begin{minipage}[b]{.45\textwidth}
  \includegraphics[width=\textwidth]{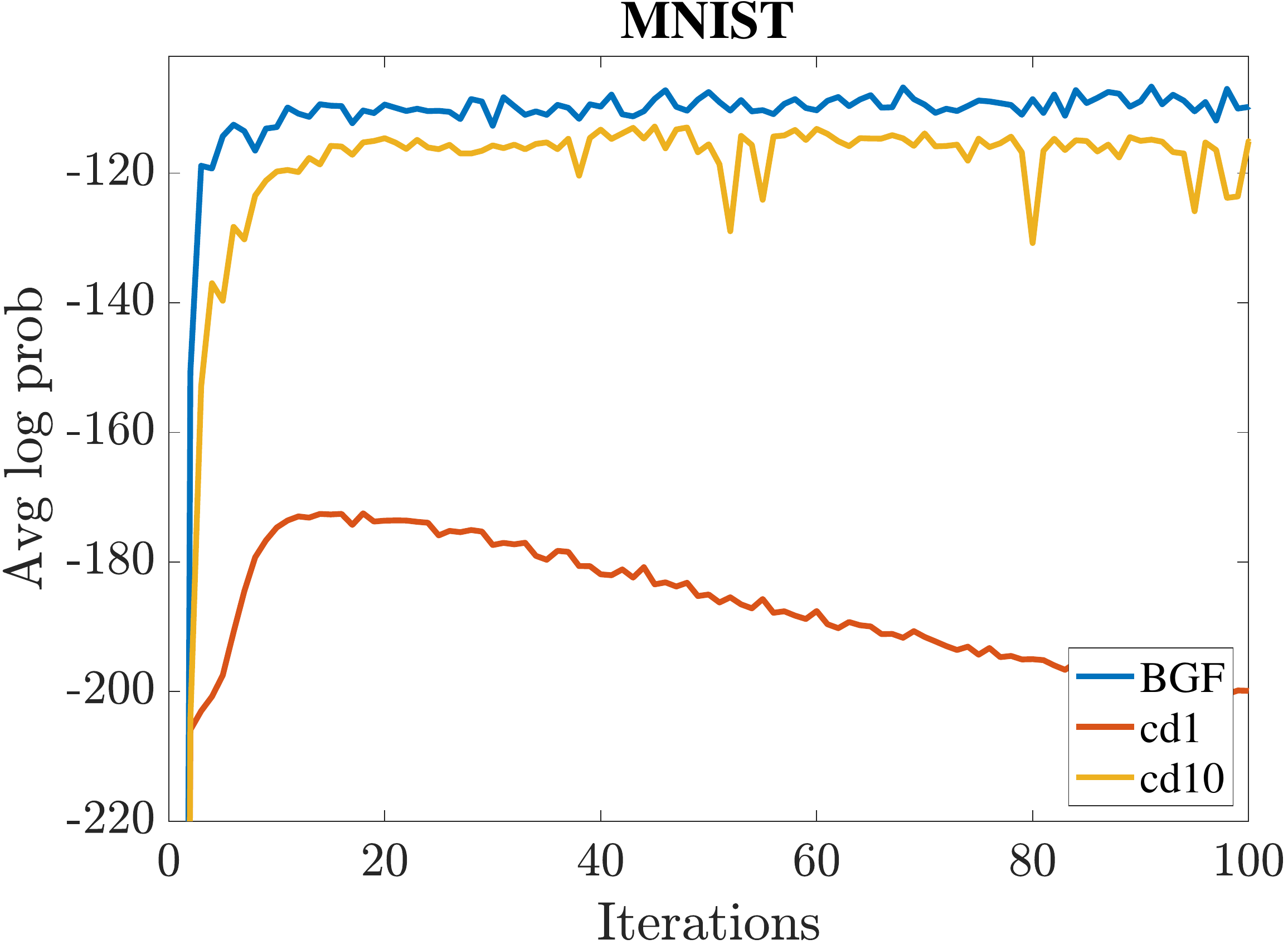}
  \end{minipage}%
  \hfill
  \begin{minipage}[b]{.5\textwidth}
  \includegraphics[width = 0.32\textwidth]{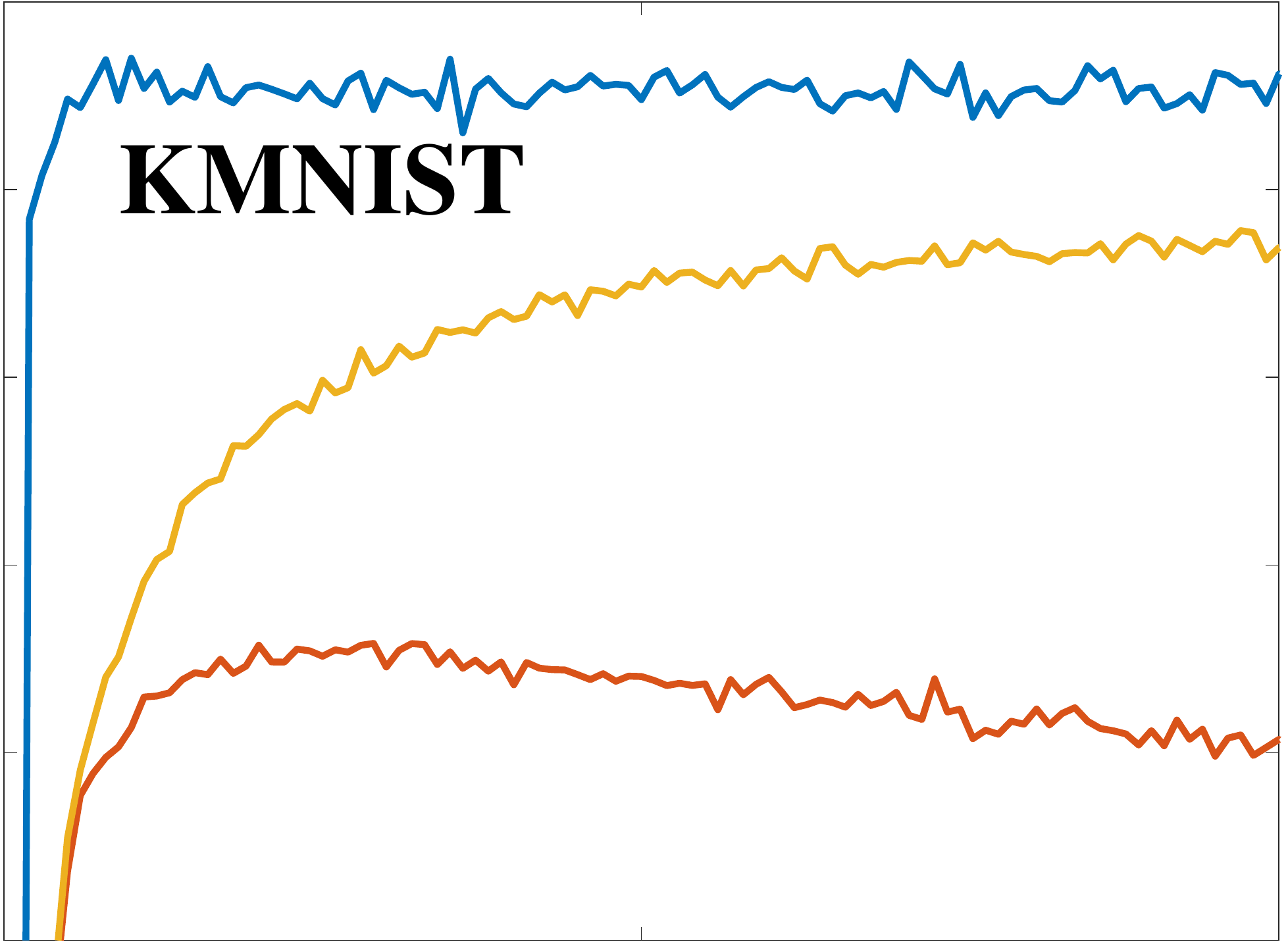}
  \includegraphics[width = 0.32\textwidth]{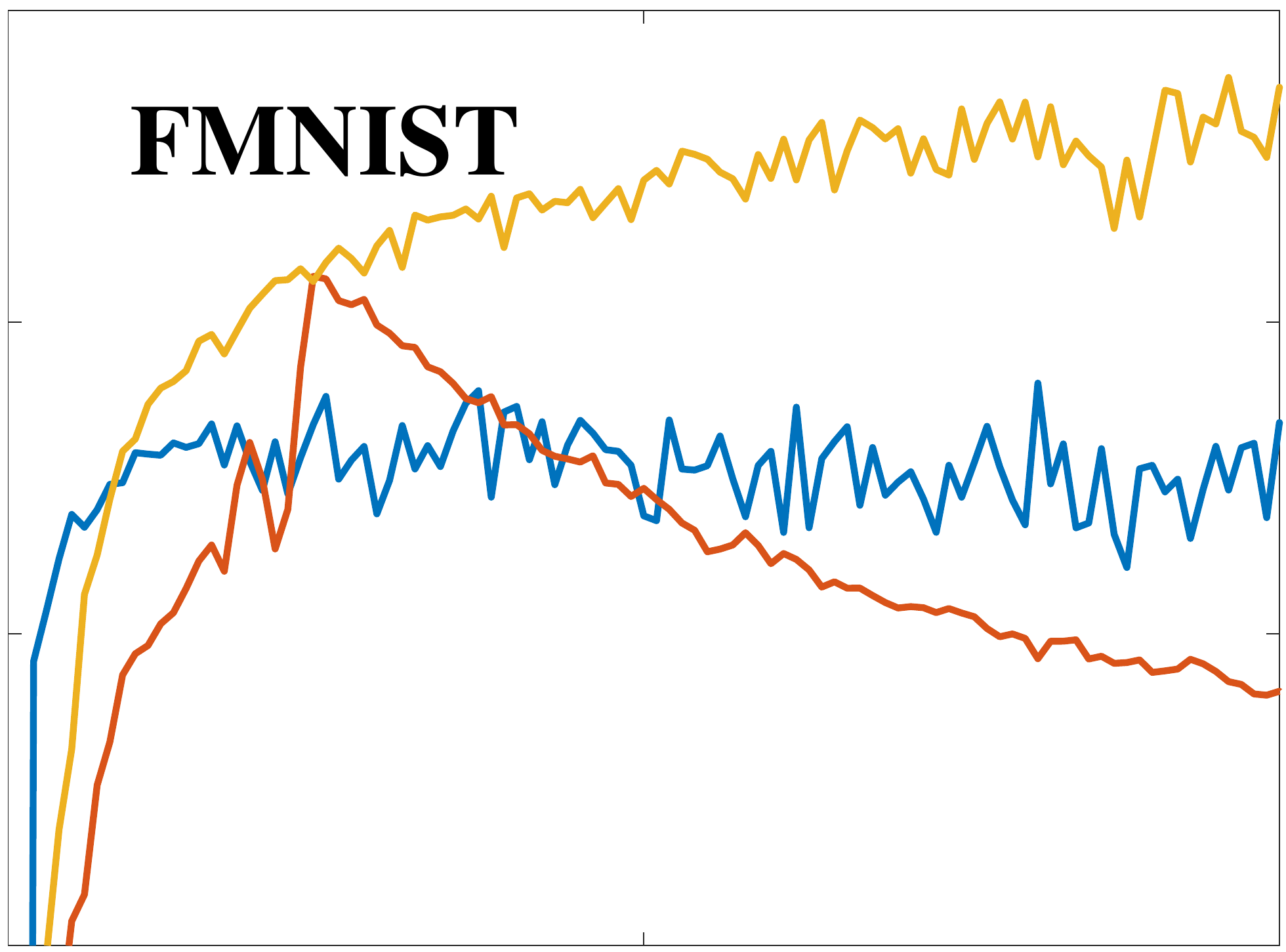} 
  \includegraphics[width = 0.32\textwidth]{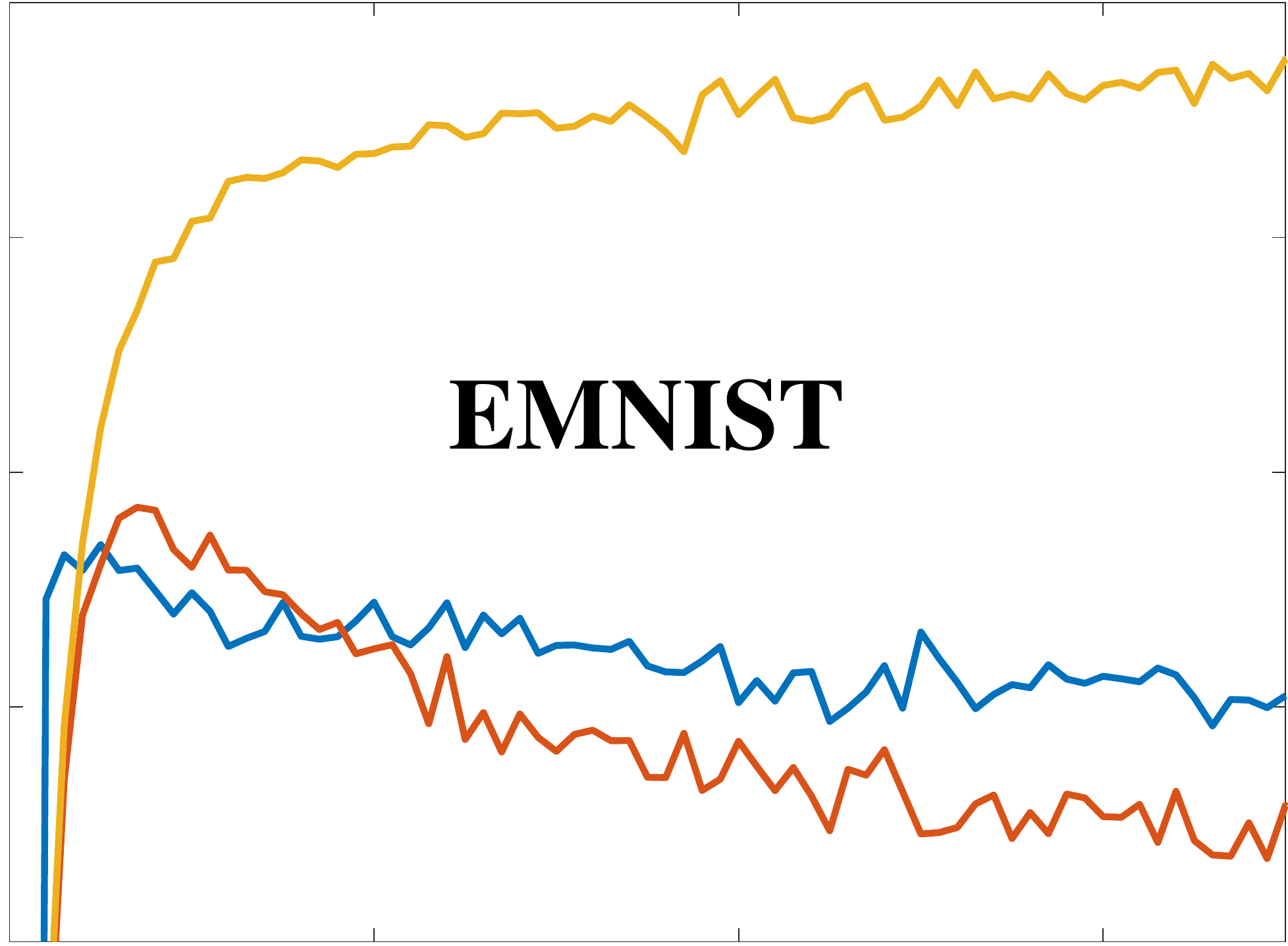}
  \end{minipage} 
  \caption{Average log probability of different models (higher is better) of conventional
algorithms (CD-1 and CD-10) and our modified algorithm used for Boltzmann 
gradient follower (BGF). The thumbnails show the general trend of other workloads.}
\label{fig:nll}
\end{figure}
Fig.~\ref{fig:nll} shows the average log probability of models
obtained using different methods 
(CD-1, CD-10 and our modified versions to Boltzmann gradient follower)
over a period of training with different data
sets. Note that the log probability is computationally intractable and
thus approximated with AIS~\cite{salakhutdinov2008quantitative} as already mentioned in 
Sec.~\ref{ssec:boltz}. 
Indeed, in some data sets (\eg CIFAR10), the same AIS mechanism
fails altogether to produce a finite estimate on the partition function.
The result therefore should not be read with much precision outside the general trend.
We show a few example data sets.

As we can see, in general, trajectories of log probability increase over time,
and often quite substantially. This means the trained models approximate
the probability distribution of training data better over time. The exact
trajectory, however, is highly variable even under common practices of CD-$k$
with different $k$ values. Our modified algorithm is understandably producing
its own trajectory. Compared to CD-10, the difference in trajectory is often
less pronounced than that resulting from choosing a small $k$ for
expediency. In addition to the uncertainty in the log probability estimate itself,
keep in mind that beyond a certain point, one could argue that better log probability
is simply a result of overfitting. Thus we also show classification errors in
Table~\ref{table:error}. Note that for the first 6 tests, the
tool by default reports only two digits of significance, which masks any
difference between using cd-$k$ and BGF. We artificially increased
the reported precision. We can see that although the results are clearly
different, both methods provide essentially the same accuracy.

\aptLtoX{\begin{table}
\caption{Test accuracy obtained by different types of Neural Network 
Models for each data set USING different algorithm.  The accuracy is 
rounded by "sklearn" (python library) function.}
\label{table:error}
\setlength{\tabcolsep}{4pt}
\begin{tabular}{|c|c|c|c|c|} 
\toprule
\bf {Datasets} & \thead{RBM \\ cd-10}  & \thead{DBN-DDN \\ cd-10 } & \thead{RBM \\BGF}   & \thead {DBN-DDN \\BGF} \\ 
\midrule
MNIST          &  95.9\%  &  98.9\%  &  96.3\%  & 98.8\%  \\ 
\hline
KMNIST         &   86.5\%  &  93.7\%  &  86.3\%  & 93.8\%  \\ 
\hline
FMNIST         &   86.3\%   &  90.1\%  &  87.1\%  &  89.8\%  \\
\hline
EMNIST         &   90.4\%   &   89.0\% & 90.4\%  & 89.0\% \\
\hline
CIFAR10        &  72.3\%  & - & 72.1\% &- \\
\hline
SmallNORB      &  72.3\%    & - & 72.4\% & -\\
\hline
Recommendation system MAE &  0.76    & - & 0.72 & -\\
\hline
Anomaly detection AUC  &  0.96    & - & 0.96 & -\\
\hline
\end{tabular}
\end{table}}{\begin{table}[htb]\centering
\vskip -10pt
\caption{Test accuracy obtained by different types of Neural Network 
Models for each data set USING different algorithm.  The accuracy is 
rounded by "sklearn" (python library) function.}
\label{table:error}
\resizebox{\columnwidth}{!}{
\setlength{\tabcolsep}{4pt}
\begin{tabular}{|c|c|c|c|c|} 
\hline
\bf {Datasets} & \thead{RBM \\ cd-10}  & \thead{DBN-DDN \\ cd-10 } & \thead{RBM \\BGF}   & \thead {DBN-DDN \\BGF} \\ 
\hline\hline
MNIST          &  95.9\%  &  98.9\%  &  96.3\%  & 98.8\%  \\ 
\hline
KMNIST         &   86.5\%  &  93.7\%  &  86.3\%  & 93.8\%  \\ 
\hline
FMNIST         &   86.3\%   &  90.1\%  &  87.1\%  &  89.8\%  \\
\hline
EMNIST         &   90.4\%   &   89.0\% & 90.4\%  & 89.0\% \\
\hline
CIFAR10        &  72.3\%  & - & 72.1\% &- \\
\hline
SmallNORB      &  72.3\%    & - & 72.4\% & -\\
\hline
Recommendation system MAE &  0.76    & - & 0.72 & -\\
\hline
Anomaly detection AUC  &  0.96    & - & 0.96 & -\\
\hline
\end{tabular}}
\end{table}}

Overall, the main takeaway point is this: We made small modifications to the
standard algorithm for the convenience of implementation. From a first
principle perspective, these changes are akin to choosing 
choosing CD-$k$ over the impractical maximum likelihood learning, 
or simply selecting a random $k$ value
in CD-$k$ for expediency reasons. Our empirical observations now indicate that these
changes indeed do not affect the general efficacy of the learning approach.

\subsection{Impact of noise on training}

\begin{figure}[htb]
  \centering
 \begin{minipage}[b]{.45\textwidth}
  \includegraphics[width=\textwidth]{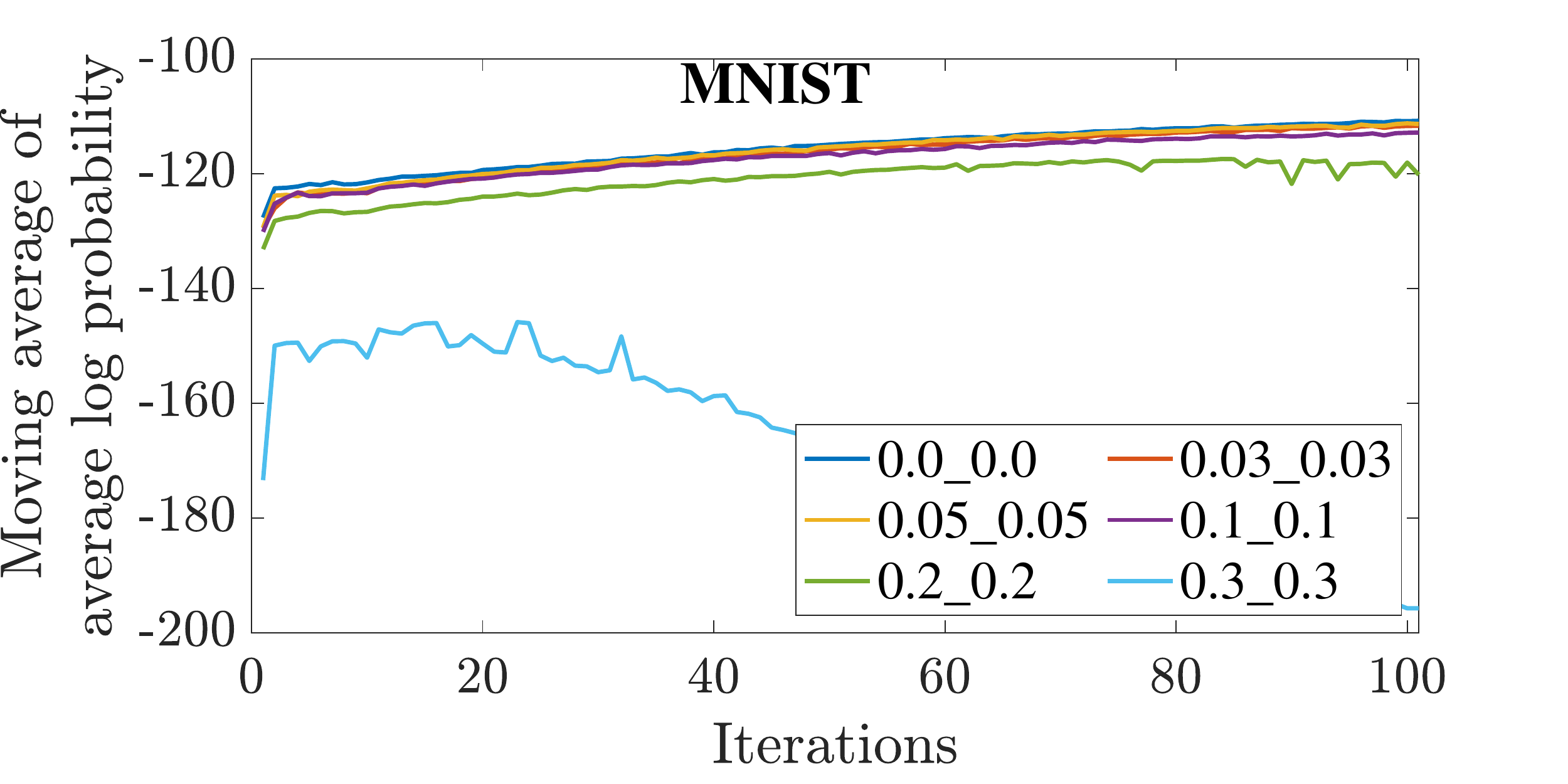}
  \end{minipage}
  %\vspace{5pt}
  \begin{minipage}[b]{.5\textwidth}
  \includegraphics[width = 0.32\textwidth]{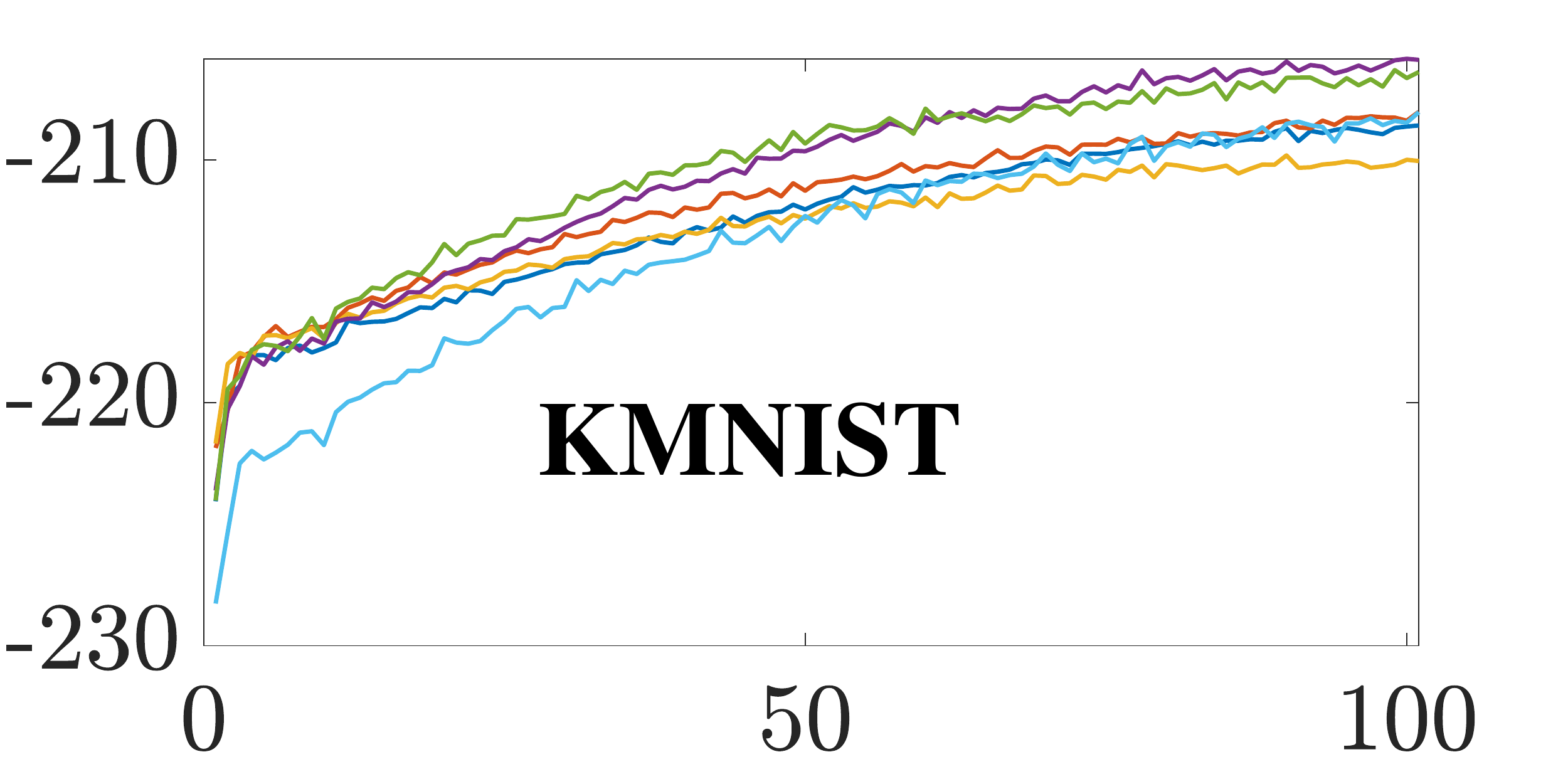}
  \includegraphics[width = 0.32\textwidth]{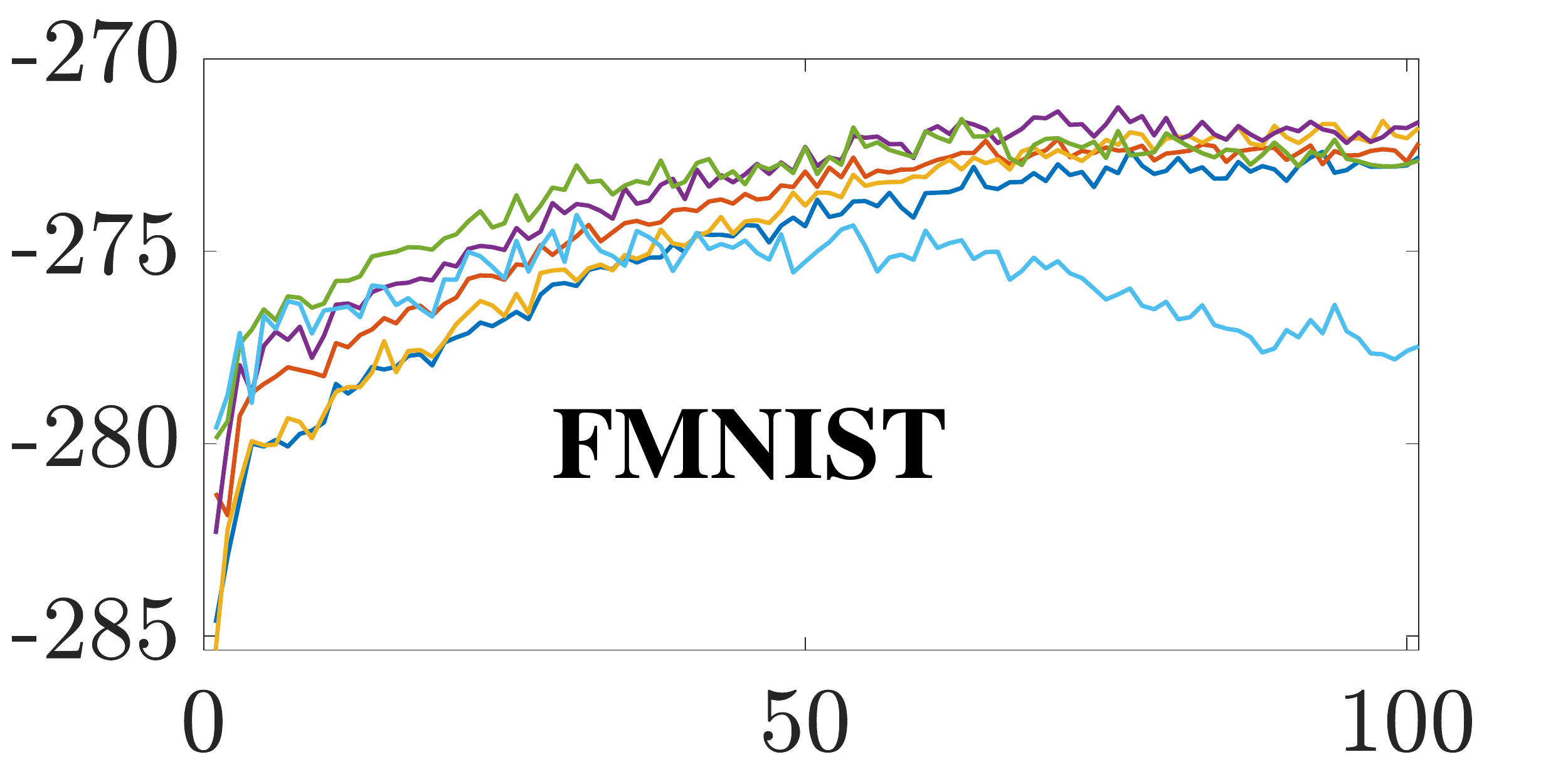}
  \includegraphics[width = 0.32\textwidth]{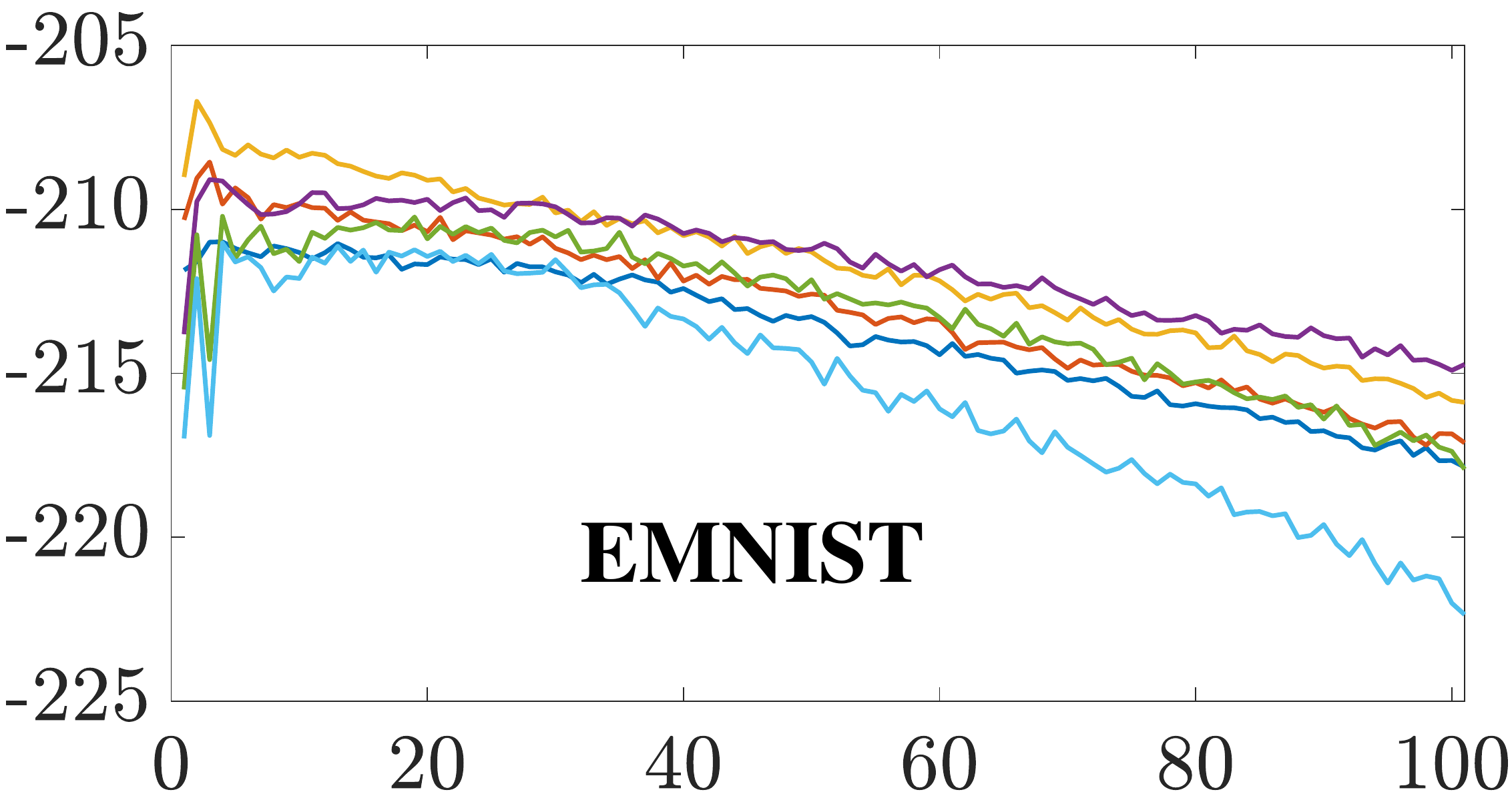}
  \end{minipage} 
  \caption{ Moving average of mean 
log probability of different models under varying amount of injected
noise and variations. 
The data are smoothed using a moving average of 10 points.}
\label{fig:smooth_noise}
\end{figure}

Finally, we look at the impact of noise and variation on solution quality.
Until this point, the results are assuming the analog hardware suffers from
no noise or variation. To simulate process variations and circuit noise, we
inject both static variation on the resistance of the coupling units and
dynamic noises at both nodes and coupling units. The noise and variation
are generated by Gaussian distribution with root mean square (RMS) values
between 3\% and 30\% for a total of 25 combinations. 
Different results are thus characterized by a pair
($RMS_{variation}, RMS_{noise}$). For clarity, in Figure~\ref{fig:smooth_noise} 
we only show the smoothed result of average log probability for 6 combinations
for a typical result (MNIST). The thumbnails show the general trend of 
other datasets.

\begin{figure}[htb]\centering
\includegraphics[width=0.4\textwidth]{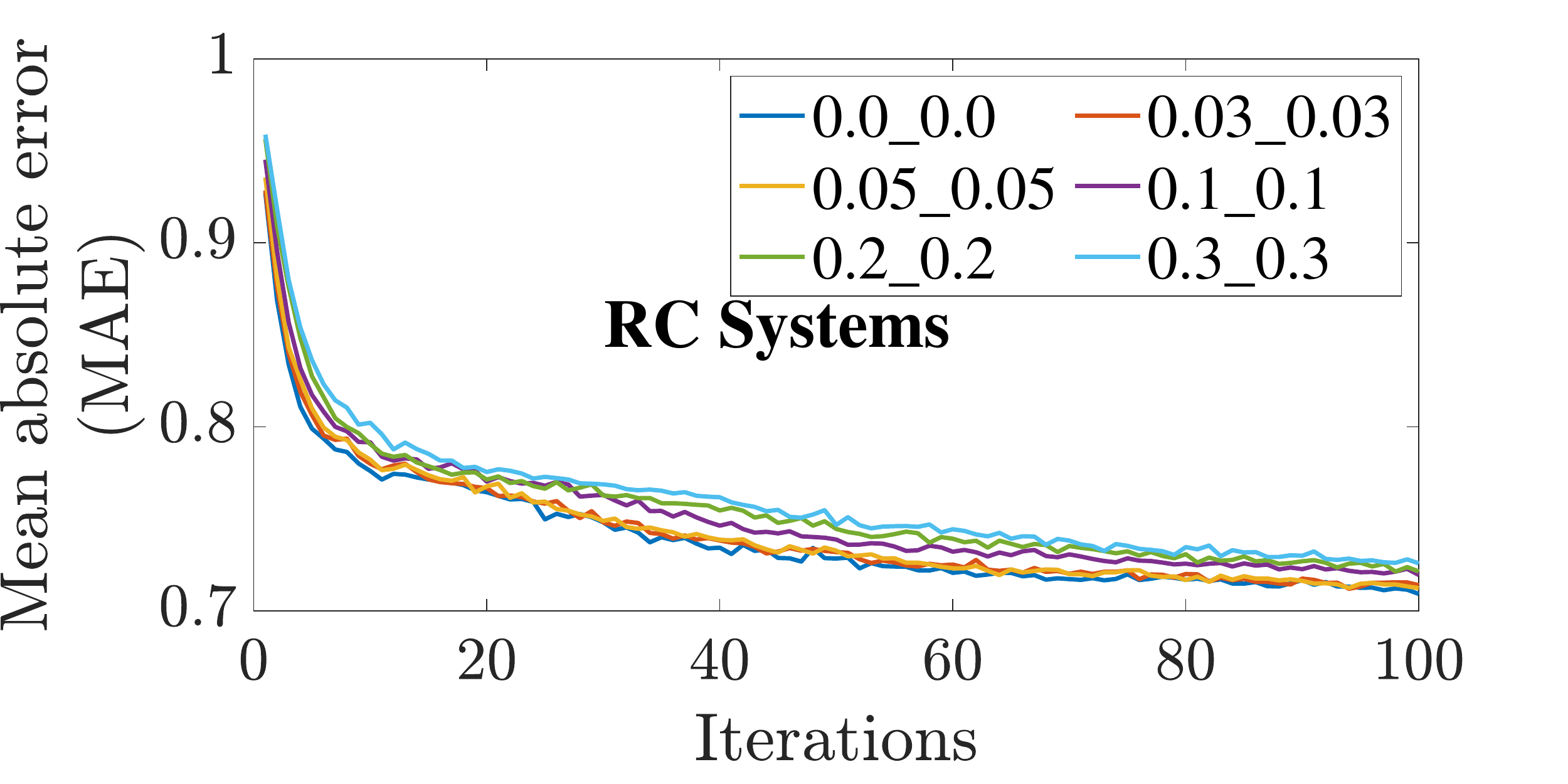}
\caption{Mean absolute error (MAE) of different models under varying amount of 
injected noise and variations. Final MAE ranges between 0.709 and 0.7258.
\label{fig:mae}}
\end{figure}

\begin{figure}[htb]\centering
\includegraphics[width=0.40\textwidth]{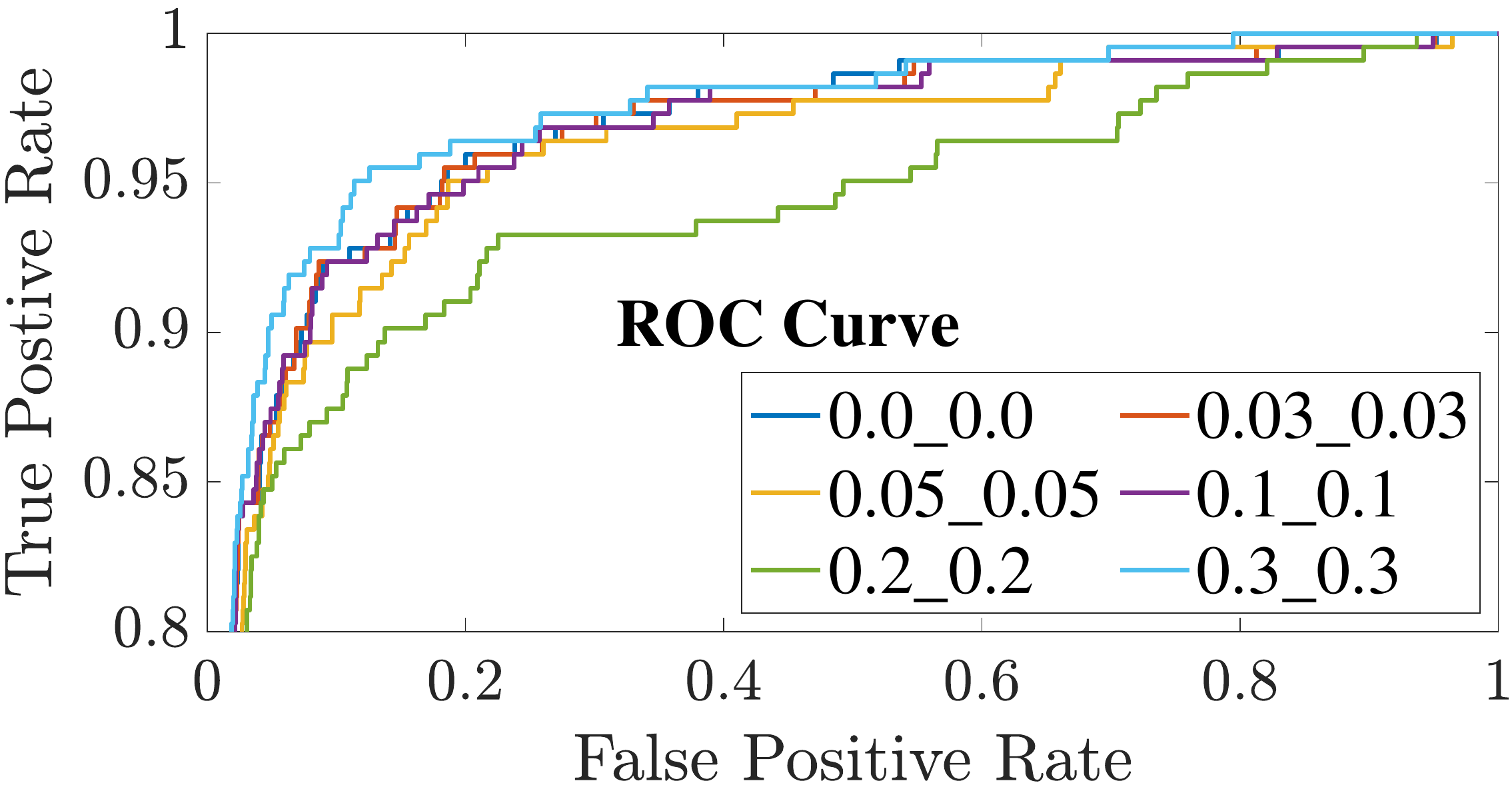}
\caption{ Roc curves of different models under varying amount of injected noise 
and  variations.  Final AUC ranges between 0.957 and 0.963.
\label{fig:fraud_detection}} \end{figure}

In general, the message is perhaps somewhat predictable: When the combination
of noise and variation is not too extreme (e.g., $<\,10\%$ each), the impact
on log probability is negligible. In many cases, the loss is smaller than that
gained by modifying the algorithm to suit our Boltzmann gradient follower. But
even for the more extreme variation and noise configurations the impact on log
probability does not appear significant. And the final inference accuracy is
essentially unchanged for all image-based benchmarks. For Recommendation system and anomaly 
detection benchmarks, the final results do show a little variation as shown
in Fig.~\ref{fig:mae} and ~\ref{fig:fraud_detection}. 

To sum, while the analog circuits are subject to noise, they can competently
perform the gradient descent function under even significant amount of noise 
without affecting the quality of the overall training process. 

\subsection{Discussions}

Overall, experimental analysis of the two architectures provides some evidence
that even without new algorithms specifically exploiting the capability of
the hardware, we can obtain substantial performance and energy benefits with a
very small additional chip footprint. On the other hand, computational substrates
such as TPU are clearly much more general-purpose. At this stage, our analysis 
does not yet
suggest that they are compelling designs per se. But we believe 
it shows that there is potential in the general direction for a number of
reasons: \begin{enumerate}[nolistsep]
     
    \item  Ising substrates are already quite useful in accelerating a 
    broad class of optimization problems. They could become even more 
    versatile and widely used in the future. Thus the incremental cost
    of additional architectural support will become lower still.
    
    \item Our designs are but an initial exploration. Further research
    could improve system's versatility and performance by, for example,
    reconfigurability and support for exploiting training 
    set parallelism.
    
    \item Energy-based models have many good qualities but are notoriously
    expensive to train. With the advent of prototype Ising machines
    and systems such as Boltzmann gradient follower, better algorithms may
    follow.

\end{enumerate}

%% file: 6conclusions.tex
\section{Conclusions}

Statistical physics has inspired a lot of effective algorithms. Physical Ising machines
are dynamical systems that embody the physics underlying popular algorithms such
as simulated annealing, RBM, and other derivative algorithms. 
As a result, an Ising machine substrate (when done right) can be used as extreme
accelerators for certain part of these physics-inspired algorithms.
In this paper, we showcased two different designs
that augment an Ising substrate with extra circuit to support RBM training.
We show that with some small changes, an Ising machine can easily serve as a Gibbs
sampler to accelerate the RBM algorithm. With more 
substantial changes, the substrate can serve as a Boltzmann sampler while
following the gradient to train RBM without any additional host
computation. Compared to an idealized TPU host significantly larger in chip area, such a 
Boltzmann gradient follower can achieve some 29x speedup and 1000x energy
savings -- though these systems are far less general-purpose than TPUs. 
With further research, we believe that hardware software co-designed
nature-based computing systems can be an important new computation modality.

%% file: 7appendix.tex
\appendix

\section{Bias of learning algorithms}
\label{sec:bias}

%\footnotesize 
%\fontsize{8.5}{9}\selectfont 
%\begin{wrapfigure}{r}{1.7in}
%    \centering
%    \vskip -10pt
%    \includegraphics[width=1.7in]{FIGS/distr1.pdf}
%    \caption{
%    Cumulative probability distribution of KL divergence of CD and BGF 
%    training result against ground truth. In other words, every point $(x,y)$ on 
%    the curve shows $y\%$ of training distribution have a final KL divergence
%    of $x$ or less from the ground truth.}
%    \label{fig:bias}
%\end{wrapfigure}

Unlike a maximum likelihood (ML) learning algorithm the contrastive 
divergence (CD) algorithm is known to be biased. This means that the fixed
points of the CD algorithm is not the same as that of the ML algorithm. 
Although the topic generated numerous publications, in practical terms
the issue is insignificant. First, the bias is shown to be small. Second,
the ultimate goal of the algorithm is to capture the training data well
enough to be useful. Nevertheless, we show empirical observations of the
bias for our modified training algorithm. For this experiment, we use the
same methodology Carreira-Perpinan and Hinton used in their original 
investigation~\cite{carreira2005contrastive}. %\begin{comment}

\begin{figure}[htb] %{r}{1.5in}
    \centering
    %\vskip -10pt
    \includegraphics[width=0.40\textwidth]{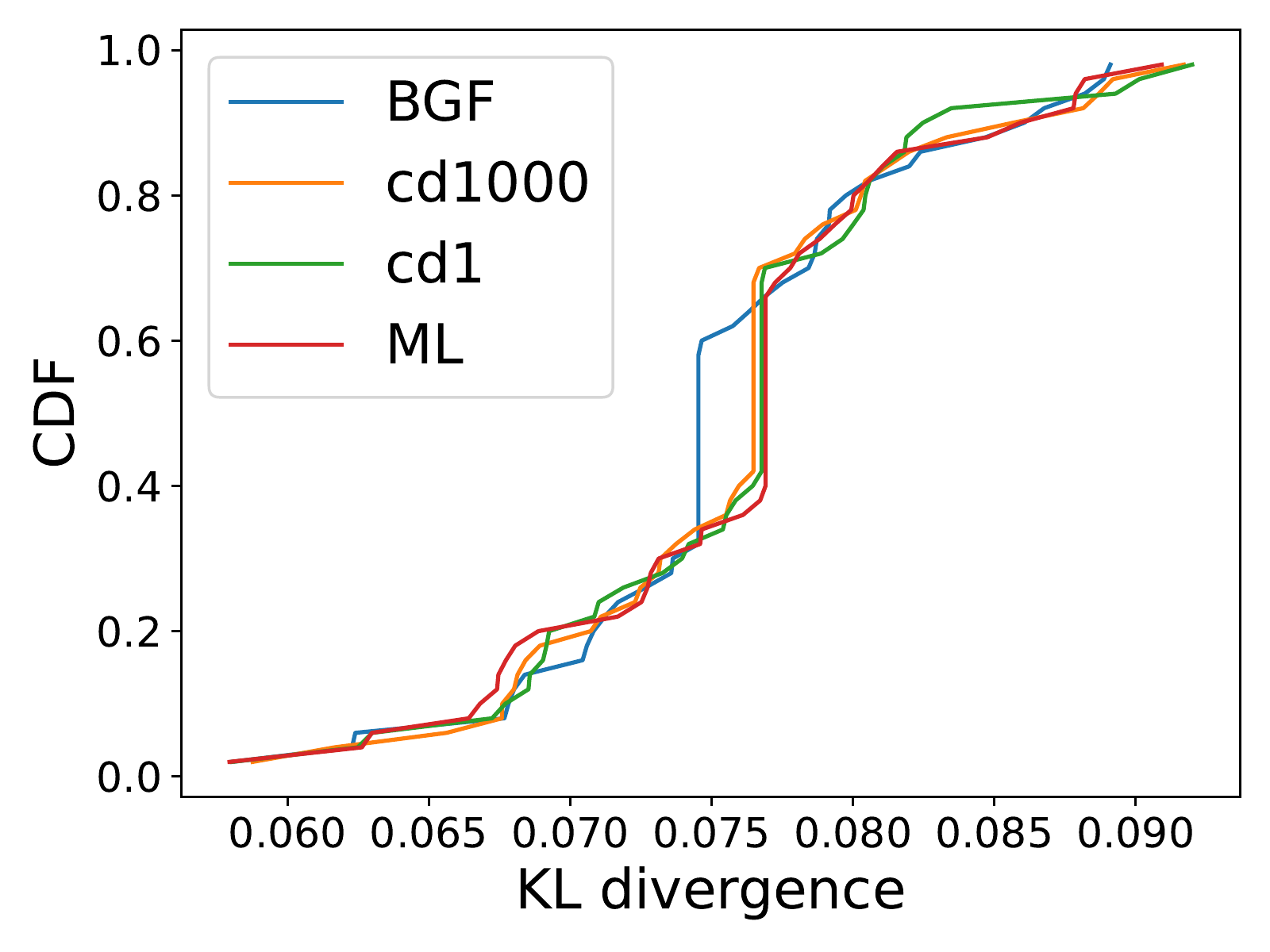}
    \caption{
    Cumulative probability distribution of KL divergence of CD and BGF 
    training result against ground truth. In other words, every point $(x,y)$ on 
    the curve shows $y\%$ of training distribution have a final KL divergence
    of $x$ or less from the ground truth.}
    \label{fig:bias}
\end{figure}

%\end{comment}

%\begin{wrapfigure}{r}{1.7in}
%\vskip -10pt
%\includegraphics[width=1.7in]{FIGS/distr2.pdf}
%\vskip -10pt
%\caption{Cumulative probability distribution of KL divergence of CD and BGF 
%    training result against ground truth. In other words, every point $(x,y)$ on 
%    the curve shows $y\%$ of training distribution have a final KL divergence
%    of $x$ or less from the ground truth.}
%\label{fig:bias}
%\end{wrapfigure}

We use a small enough system size that the 
ground truth can be obtained via enumeration. The system consists of
12 visible units and 4 hidden units, all binary. We generate 60 different
distributions of 100 training images randomly. We then execute ML, CD,
and our training algorithm (BGF) for the same 1000 iterations to
obtain the weights. Finally, we compare the resulting probability distribution
against the ground truth by measuring the KL divergence. Each run
of the algorithm produces one KL divergence measure. We repeat 400 runs
for each algorithm from different random initial conditions. The resulting
400 different measures are plotted as cumulative probability distribution shown in
Fig.\ref{fig:bias}.

We can see that all the algorithms have a fairly similar
bias characteristic. This is not not surprising since BGF is really a modified
CD algorithm. However, because of the inherent speed in exploring the phase
space, BGF can be thought of as CD-$k$ with a very large $k$. As $k\to \infty$,
CD-$k$ effectively becomes ML. As a result, we see BGF indeed offers less chance
of a larger KL divergence compared to conventional CD-$k$. Overall, the takeaway
point is clear. Our BGF does not create a problem of biased estimation. If
anything, it improves the bias characteristic of the commonly used von Neumann
algorithm.

%% file: 4circuit.tex
\section{Circuit}\label{sec:circuit}

In addition to the circuits needed to implement the baseline Ising substrate,
we need extra circuits in the nodes to make them probabilistic according to RBM
algorithms, which are summarized in Fig.~\ref{fig:sampling_high_level}. Specifically, we need a current summing circuit (Sec.~\ref{ssec:current_summation}), a sigmoid function (Sec.~\ref{ssec:sigmoid}), 
and a random number generator (Sec.~\ref{ssec:rng}). 
Finally, for the Boltzmann gradient follower architecture, the coupling unit
is discussed in (Sec.~\ref{ssec:cu}).

\subsection{Current summation}\label{ssec:current_summation}
\begin{figure}[hbt]\centering
\includegraphics[width=0.45\textwidth]{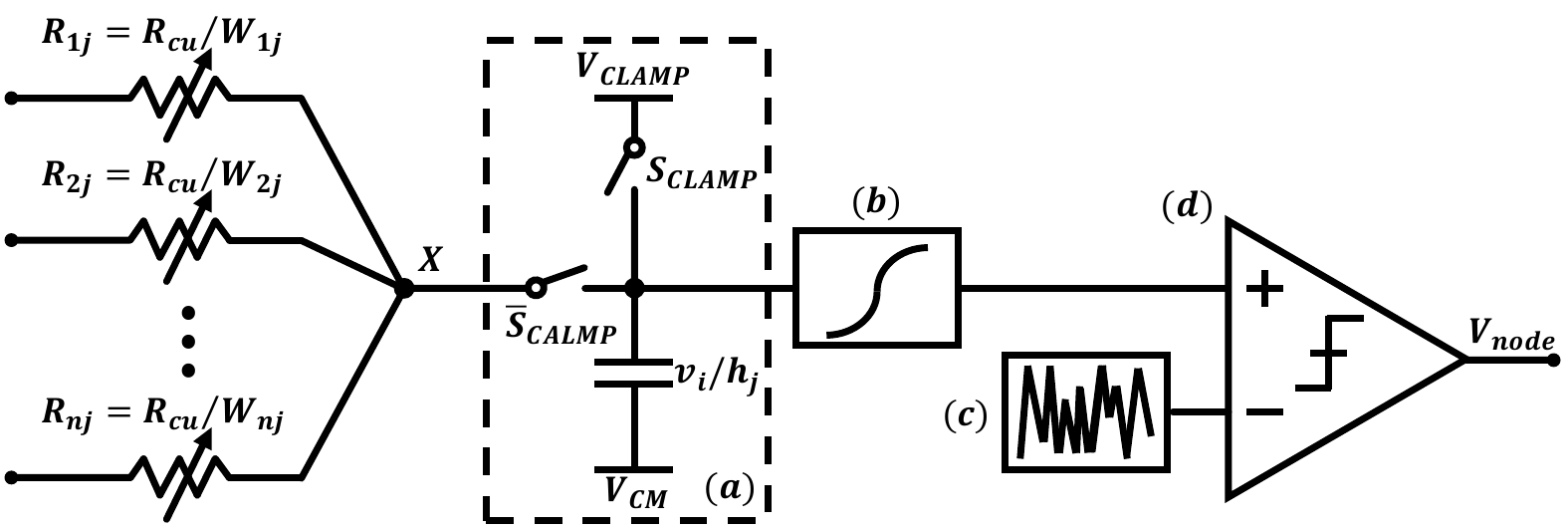}
\caption{Component-level diagram for a node. Detailed implementations of (b), (c) and (d) are depicted in Fig.~\ref{fig:sampling_low_level}. } \label{fig:sampling_high_level}
\end{figure}

The summation process illustrated in Fig.~\ref{fig:sampling_high_level}(a) consists of two distinct operating phases referred to as $S_{CLAMP}$. During the clamp phase ($S_{CLAMP}$ is closed), the visible nodes($v_i$) are clamped to training samples, while the hidden nodes($h_j$) are clamped to $V_{CLAMP}$ for initiation and reset purposes. During the free-running phase (when $S_{CLAMP}$ is open), the current flowing from $R_{ij}$ begins to integrate on the capacitor, resulting in a gradual voltage shift. The $dc$ values of nodes are set by bias one terminal of the capacitor to $V_{CM}=V_{dd}/2$.

\subsection{Sigmoid unit}\label{ssec:sigmoid}

A Sigmoid function, which is formulated as $S(x) = \frac{1}{1 + e ^{-c_1 \cdot (x - c_2)}}$, commonly serves as an activation function that introduces non-linearity in neural networks or converts the output of a linear model to probability for certain applications. The two hyper-parameters, $c_1$ and $c_2$, allow fine-tuning of the Sigmoid curve, adapting it to diverse situations. This same functionality can be achieved using a differential to single-ended amplifier, as illustrated in Fig.~\ref{fig:sampling_low_level}(a). In this configuration, the amplifier's gain is intentionally set to a lower level, making the transfer function closely resemble a Sigmoid shape. An advantage of this approach lies in its flexibility, as the shape of the transfer function can be adjusted through a current bias control. By choosing different values for $V_{hp}$, the gain of the amplifier can be modified accordingly, which is similar to tuning the hyper-parameters in a Sigmoid function. 
%It is important to note that the inverter's transfer function exhibits a vertically flipped Sigmoid shape, which does not impact the non-linearity introduced into the learning model, while it can be rectified by using a subsequent inverting comparator shown in Fig.~\ref{fig:sampling_low_level}(c).

%The first two issues can be alleviated by attaching resistors at both the input and the output. The one at the input is used to convert current to voltage while the one at the output works as a load to attenuate an undesirably high gain of the inverter. The last issue is automatically solved when we connect the sigmoid unit to the current summation unit, which will be discuss in \ref{ssec:CC}. 

%The actual circuit implementation is as shown in Figure \ref{fig:sampling_low_level} (a), where we use two transmission gates ($TG_1$ and $TG_2$) to play the role of the resistors discussed above. The simulated I-V characteristic curve of this circuit as well it's sigmoid-approximation are plotted in Figure \ref{fig:sigmoid_IV}. 

\subsection{Random number generator}\label{ssec:rng}
%Thermal noise from electronic devices can be utilized to generate randomness. The circuit shown in Fig.~\ref{fig:sampling_low_level}(b) 
%is one kind of random number generator (RNG) and its operation is as follows.
%When $\phi$ is low, nodes A and B are pulled up to 
%$V_{dd}$, which is a metastability point. When $\phi$ goes high, the circuit
%enters into the evaluation phase: both nodes discharge towards the switching 
%point of the inverter, then the large gain at this point pulls one of the nodes to $V_{dd}$, while the other is pulled down to ground, depends on the 
%sign of the differential thermal noise. The RNG circuit itself produces a binary random sequence (i.e., the output is either $V_{dd}$ or $0$), which is not directly suitable to achieve probabilistic sampling at the output of the sigmoid function. However, the binary random sequence at the %output of the RNG is readily converted to a white-noise uniformly distributed from $0$ to $V_{dd}$, by applying an RC low-pass filter as shown in Fig.~\ref{fig:sampling_high_level}(d). The resulting white-noise is then compared against the output of the sigmoid function in a standard %dynamic comparator \cite{baker2019cmos} (shown in Fig.~\ref{fig:sampling_low_level}(c)) to achieve probabilistic sampling. 

Regularly, noise generators and chaotic-signal generators are two approaches to producing random numbers in the most of literature. In this study, we leverage thermal noise in the diode to create a random number generator, as shown in Fig.~\ref{fig:sampling_low_level}(b). Two diodes $D_1$ and $D_2$ perform as noise source generators as the input of variable-gain noise amplifiers. The amplifier is biased at $V_{CM}=V_{dd}/2$ so finally the noise is amplified to a random number distributed from ($V_{CM}-A\cdot{V_{noise}}$) to ($V_{CM}+A\cdot{V_{noise}}$). The generated noise signal is subsequently applied to the input of a standard dynamic comparator \cite{baker2019cmos}, compared to the output of the Sigmoid unit, thus enabling probabilistic sampling. The output buffer(typically a latch according to \cite{baker2019cmos}) servers to digitize the output voltage and ensure sufficient drive capability. Multiple devices can be used as a diode such as gate-drain shorted MOSFET or base-collector shorted BJT to reduce the chip area.

\subsection{Coupling unit for Boltzmann gradient follower}\label{ssec:cu}

\aptLtoX{\begin{figure}
    \centering
  \centering
  \includegraphics[width=\textwidth]{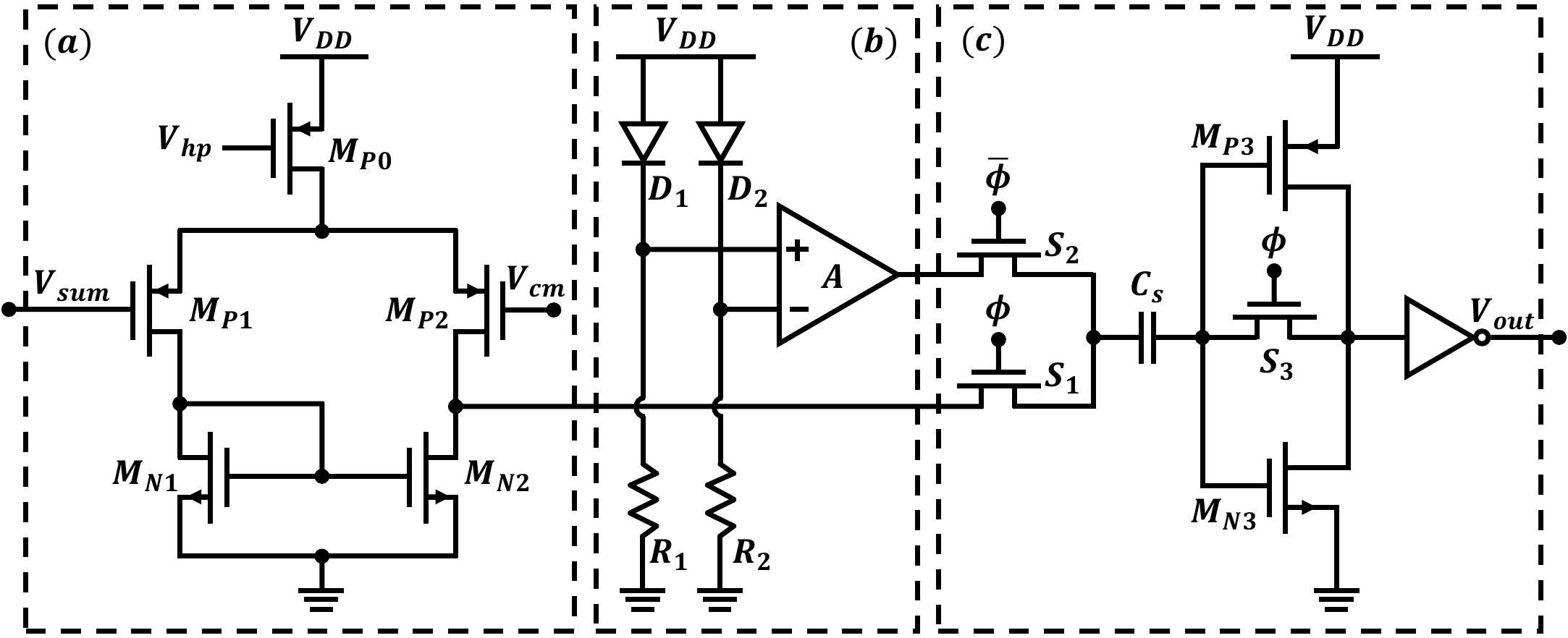}
  \caption{(a) Sigmoid unit; (b) Random noise generator; (c) Dynamic comparator}
  \label{fig:sampling_low_level}
\end{figure}
\begin{figure}
  \centering
    \includegraphics[width=\textwidth]{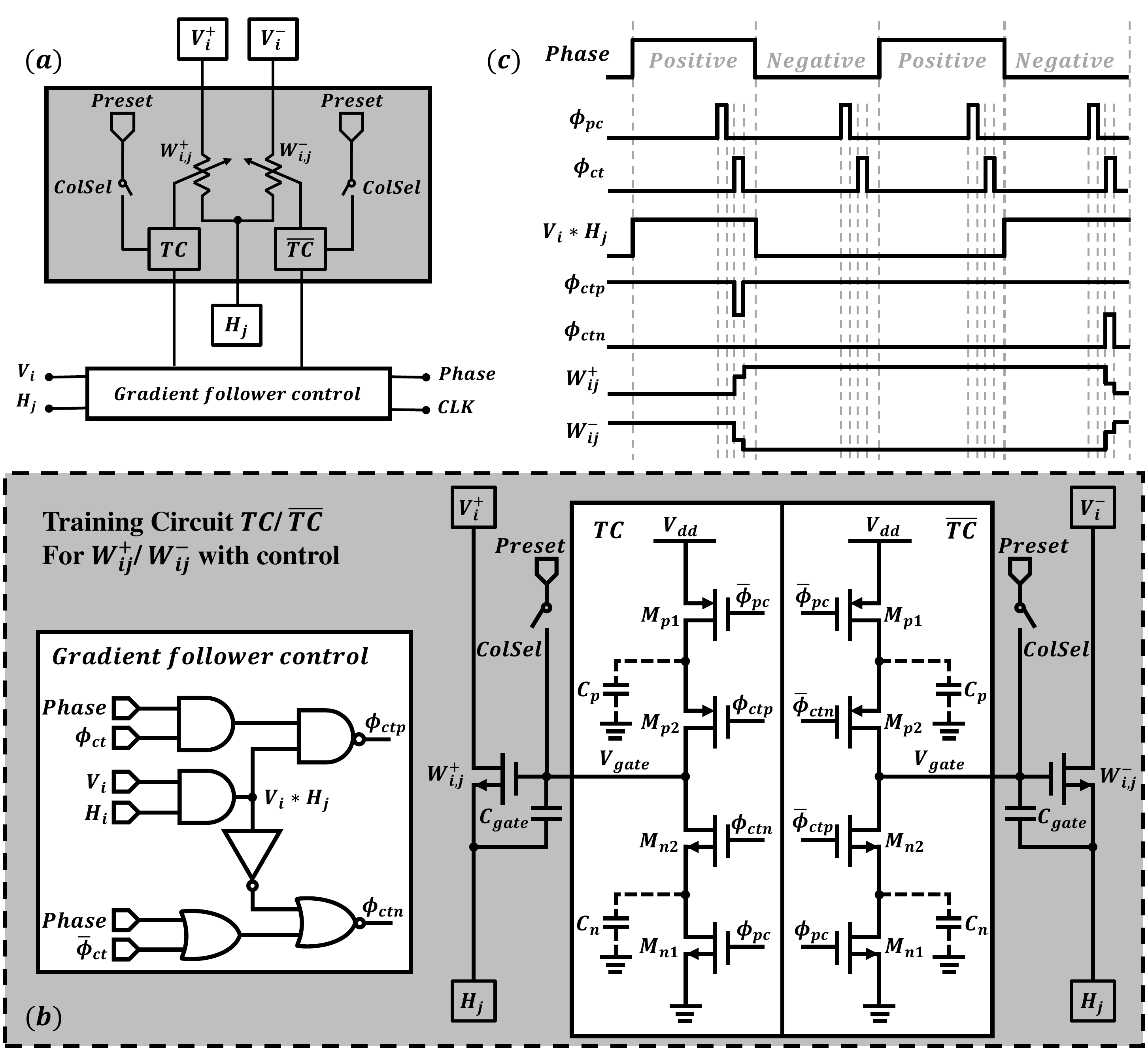}
    \caption{(a). Component diagram for coupling unit with training circuit. (b). Schematic details. (c). Timing diagram for one of the possible system state.}
    \label{fig:CU_scheme}
\end{figure}}{\begin{figure}[htb]
    \centering
\begin{minipage}{0.45\textwidth}
  \centering
  \includegraphics[width=\textwidth]{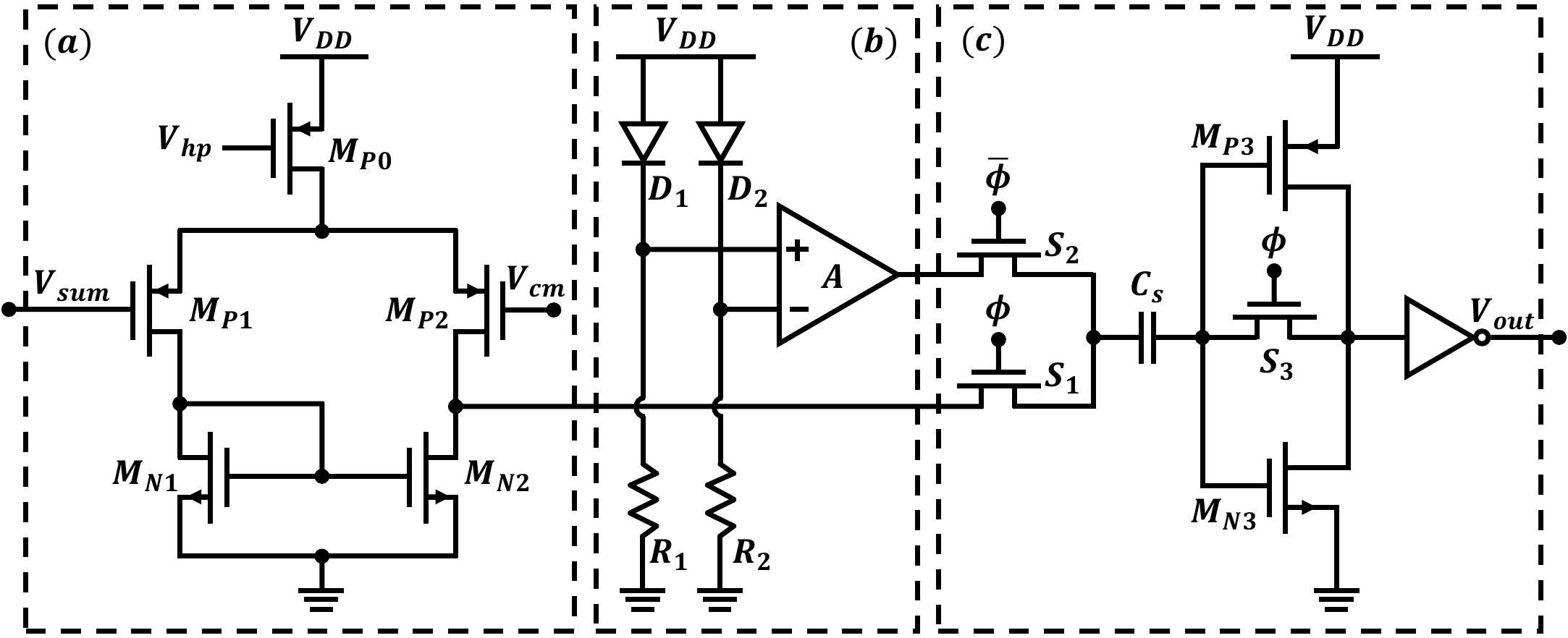}
  \caption{(a) Sigmoid unit; (b) Random noise generator; (c) Dynamic comparator}
  \label{fig:sampling_low_level}
\end{minipage}
\begin{minipage}{0.45\textwidth}
  \centering
    \includegraphics[width=\textwidth]{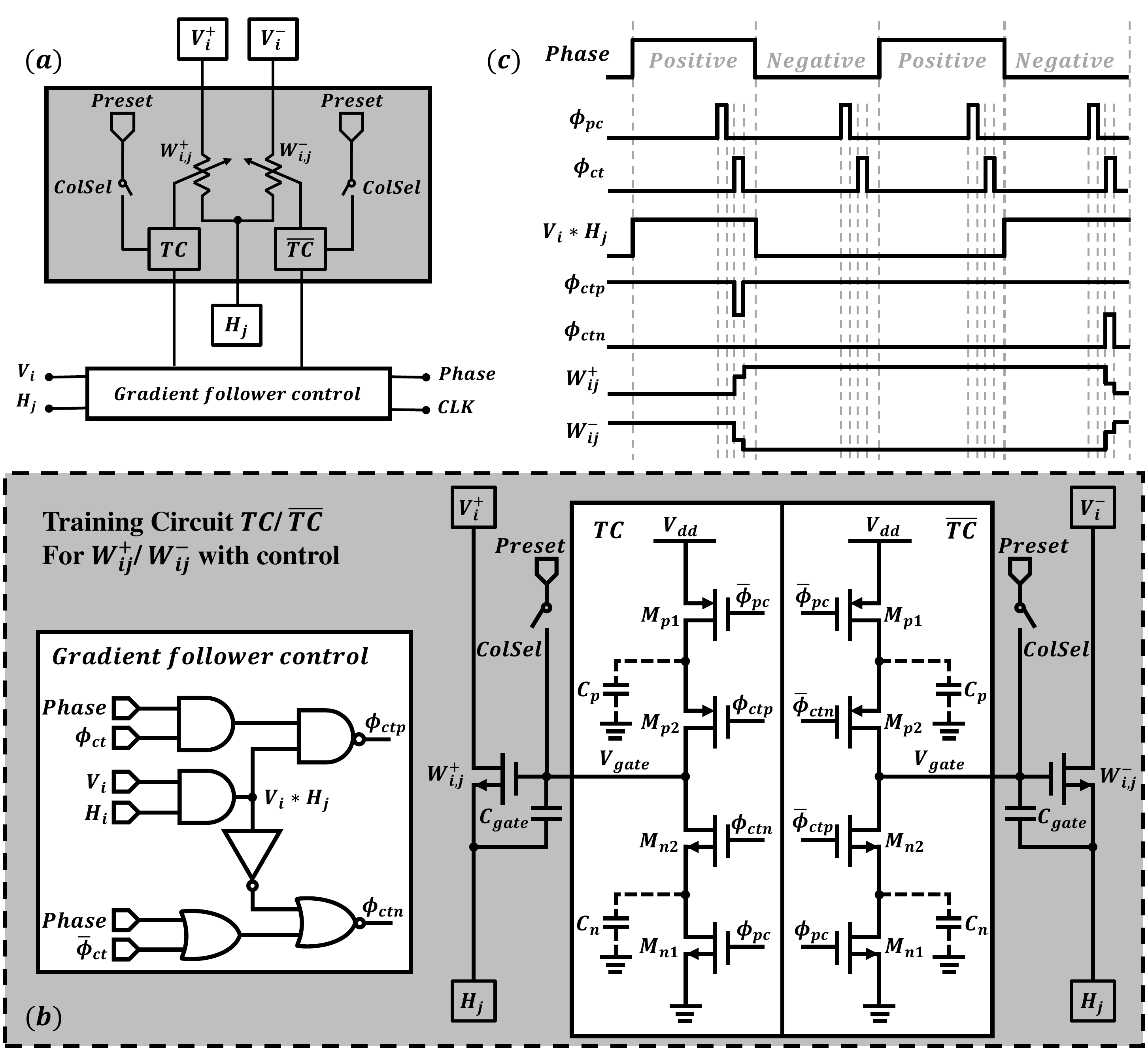}
    \caption{(a). Component diagram for coupling unit with training circuit. (b). Schematic details. (c). Timing diagram for one of the possible system state.}
    \label{fig:CU_scheme}
 \end{minipage}%
\end{figure}}

For each coupling unit representing a coupling parameter ($W_{i,j}$), a training circuit($TC$/$\overline{TC}$) is introduced to allow the Boltzmann gradient follower algorithm to adjust $W_{i,j}$ as shown in Fig.~\ref{fig:CU_scheme}(a). This adjustment is achieved through a charge redistribution-based CMOS charge pump circuit depicted in Fig.~\ref{fig:CU_scheme}(b). The charge pump consists of two pre-charging transistors ($M_{p1}$ and $M_{n1}$) and two charge transfer transistors ($M_{p2}$ and $M_{n2}$), driven by two non-overlapping control signals ($\phi_{pc}$ and $\phi_{ct}$). $\phi_{ct}$ is regulated by phase control to differentiate between the increment and decrement of $W_{i,j}$ during different phases (positive/negative). $C_p$ and $C_n$ are parasitic capacitance at those internal nodes. Note that we use a complementary phase control for charge transfer transistors to tune $TC$ and $\overline{TC}$, enabling the update of $W^+_{ij}$ and $W^-_{ij}$ in opposite directions.

The operation of the coupler, including training circuit is as follows. To be explicit, we only discuss the increment/decrement of $W^+_{i,j}$ here. During the pre-charge phase ($\phi_{pc}$ is high), $C_p$ will be precharged to $V_{dd}$, resulting in a charge storage of $C_pV_{dd}$ on $C_p$. Concurrently, $C_n$ is discharged to ground. Subsequently, during the charge transfer phase (when $\phi_{pc}$ is low),

\textbf{Positive sample phase($Phase = 1$):} 
If $V_i*H_j=1$, $M_{p2}$ will be activated, leading to charges redistributed from $C_p$ to $C_{gate}$, and hence increase $V_{gate}$($W^+_{i,j}$). 
If $V_i*H_j=0$, $M_{p2}$ remains off, thereby preserving the value of $W_{i,j}$.
The discharging path $M_{n2}$ is always turned off during the whole positive sample phase to ensure that $W^+_{i,j}$ will not be decremented.

\textbf{Negative sample phase($Phase = 0$):} 
If $V_i*H_j=1$, $M_{n2}$ will be activated to redistribute charges on $C_{gate}$ with $C_n$,  resulting in a decrease in $V_{gate}$ ($W^+_{i,j}$).
If $V_i*H_j=0$, $M_{n2}$ is kept off, thus maintaining the value of $W^+_{i,j}$.
$W^+_{i,j}$ will not be incremented because the charging path $M_{p2}$ is always turned off during the whole negative sample phase.

The timing diagram illustrates one of the possible system states for the coupler, as depicted in Fig.~\ref{fig:CU_scheme}(c). Due to the non-overlapping control signals, there will be no current path directly connecting $C_{gate}$ to $V_{dd}$ or ground. So the charges transferred onto $C_{gate}$ per cycle are determined by the ratio between by $C_p$, $C_n$ and $C_{gate}$. By circuit design, this charge transfer can be accurately controlled to achieve a step size of only a small number of electrons.